\documentclass[twocolumn,twocolappendix]{aastex701}

\usepackage{times}

\usepackage[T1]{fontenc}

\usepackage[utf8]{inputenc}
\usepackage{CJKutf8}
\newcommand{\cjk}[1]{
  \begin{CJK}{UTF8}{gbsn}#1
\end{CJK}}

\usepackage{inconsolata}
\usepackage{hyperref}
\usepackage{url}
\usepackage{booktabs}
\usepackage{graphicx}
\usepackage{amsmath}
\usepackage{wrapfig}
\usepackage{lipsum}
\usepackage{tikz}
\usepackage{subcaption}
\usepackage[dvipsnames]{xcolor}
\usepackage{pgfplots}
\usetikzlibrary{positioning, shapes.multipart, backgrounds, fit, math}
\usepackage{soul}
\usepackage{color}
\usepackage[all]{nowidow}
\usepackage{needspace}
\usepackage{float}
\usepackage{lipsum}

\hypersetup{linkcolor=OliveGreen,citecolor=MidnightBlue,filecolor=cyan,urlcolor=BrickRed}

\newcommand{\teff}{\textrm{T}_{\rm eff}}
\newcommand{\logg}{\textrm{log}\,\textit{g}}
\newcommand{\feh}{\textrm{[Fe/H]}}

\begin{document}

\title{Homogeneous Stellar Parameters from Heterogeneous Spectra with
Deep Learning}%

\correspondingauthor{Jeff Shen}

\author[orcid=0000-0001-6662-7306]{Jeff Shen}
\affiliation{Department of Astrophysical Sciences, Princeton
University, 4 Ivy Ln, Princeton, NJ 08540, USA}
\email[show]{shenjeff@princeton.edu}

\author[orcid=0000-0003-2573-9832]{Joshua S. Speagle (\cjk{沈佳士})}
\affiliation{Department of Statistical Sciences, University of
  Toronto, 9th Floor, Ontario Power Building, 700 University Avenue,
Toronto, ON M5G 1Z5, Canada}
\affiliation{David A. Dunlap Department of Astronomy \& Astrophysics,
University of Toronto, 50 St George Street, Toronto, ON M5S 3H4, Canada}
\affiliation{Dunlap Institute for Astronomy \& Astrophysics,
University of Toronto, 50 St George Street, Toronto, ON M5S 3H4, Canada}
\affiliation{Data Sciences Institute, University of Toronto, 17th
  Floor, Ontario Power Building, 700 University Avenue, Toronto, ON M5G
1Z5, Canada}
\email{j.speagle@utoronto.ca}

\author[orcid=0000-0002-1068-160X]{Shirley Ho}
\affiliation{Center for Computational Astrophysics, Flatiron
Institute, 162 Fifth Avenue, New York, NY 10010, USA}
\affiliation{Department of Astrophysical Sciences, Princeton
University, 4 Ivy Ln, Princeton, NJ 08540, USA}
\affiliation{Department of Physics \& Center for Data Science, New
York University, 726 Broadway, New York, NY 10003, USA}
\email{shirleyho@flatironinstitute.org}

\begin{abstract}

  Large-scale spectroscopic surveys have collectively observed
  millions of stars across the Milky Way, but each derives stellar
  labels using independent pipelines with distinct modelling
  assumptions, introducing systematic offsets that obscure signals in
  chemical space and hinder large-scale Galactic archaeology.
  We present a unified deep-learning framework that delivers
  atmospheric parameters, chemical abundances for 20 elements,
  distances, and ages---all on a single, self-consistent scale---for
  an arbitrary number of spectroscopic surveys simultaneously.
  Our approach uses a Transformer model that ingests spectra of
  arbitrary wavelength range and resolution, trained end-to-end as a
  single model across all surveys, eliminating the need for post-hoc
  recalibration.
  We apply this framework to spectra from APOGEE DR17, GALAH DR3,
  DESI DR1, and \textit{Gaia} RVS DR3, spanning resolutions from $R
  \sim 2{,}000$ to $28{,}000$ and wavelengths from the optical to the
  near-infrared.
  On high-resolution APOGEE spectra the model achieves precisions of
  $18$~K in $\teff$, $0.04$~dex in $\logg$, $0.015$~dex in $\feh$,
  and ${<}\,0.03$~dex across all abundances; on lower-resolution DESI
  spectra, typical precisions are $51$~K, $0.09$~dex, $0.04$~dex, and
  ${\sim}\,0.06$~dex, respectively.
  Cross-survey comparisons demonstrate that labels for the same stars
  observed by different surveys are consistent within model
  uncertainties; we further validate against external distance
  catalogs and open cluster metallicities and ages.
  The resulting homogeneous catalog enables Galactic archaeology at
  unprecedented scale and consistency, and the framework is readily
  extensible to forthcoming spectroscopic surveys such as SDSS-V,
  WEAVE, and 4MOST.
  The catalog is publicly available at
  \url{https://doi.org/10.5281/zenodo.19830515}.

\end{abstract}

\section{Introduction}

We are in the golden age of large-scale stellar spectroscopy.
Ground-based surveys such as the Sloan Digital Sky Survey
\citep[SDSS;][]{York2000}, its successor programs SEGUE
\citep{Yanny2009}, BOSS \citep{Dawson2013}, and eBOSS
\citep{Dawson2016}, the Apache Point Observatory Galactic Evolution
Experiment \citep[APOGEE;][]{Majewski2017}, the GALactic Archaeology
with HERMES survey \citep[GALAH;][]{buder_galah_2021}, and the Dark
Energy Spectroscopic Instrument survey \citep[DESI;][]{DESIDR1} have
measured spectra for millions of stars across the Milky Way with
exquisite precision.
Space-based missions such as the \emph{Gaia} Radial Velocity
Spectrometer \citep[RVS;][]{GaiaRVS} add millions more.
These datasets, combined with precise astrometry and photometry from
\emph{Gaia} \citep{GaiaDR3}, provide an unprecedented view of the
chemo-dynamical structure of the Galaxy.

Chemical abundances measured from stellar spectra are particularly
powerful for Galactic archaeology.
Because stars largely preserve the chemical composition of their
birth gas cloud, abundances serve as a fossil record of star
formation and enrichment history across cosmic time \citep{Freeman2002}.
The multi-dimensional abundance space---spanning $\alpha$-elements,
iron-peak elements, and neutron-capture elements---encodes distinct
nucleosynthetic channels and timescales, and together with precise
ages from, e.g., asteroseismology \citep{Pinsonneault2018}, enable
the identification of substructure \citep[e.g.,][]{Helmi2018,
Myeong2018, Berni2025}, chemically distinct stellar populations
\citep[e.g.,][]{Bensby2014, Hayden2015}, and the imprints of radial
migration \citep[e.g.,][]{Frankel2018, Frankel2019, Frankel2020,
Lian2022radialmigration}.

Despite the wealth of available data, progress in Galactic
archaeology is hampered by the fact that each survey delivers its own
stellar labels ($\teff$, $\logg$, $\feh$, individual elemental
abundances, distances, and ages) using independent pipelines, each
with its own modelling assumptions and systematics.
As a result, even for stars observed by multiple surveys, differences
in labels from different pipelines can disagree at a level comparable
to or larger than the astrophysical signals of interest.
These inter-survey offsets can wipe out any signal of chemical trends
across time and space, complicate cross-survey comparisons, and make
it difficult to fully leverage the wealth of complementary datasets
for Galactic archaeology.
Ideally, we would have a single, unified analysis pipeline for all
available spectra so that their derived labels are all on a common
scale---but to date there has been difficulty in achieving this due
to the widely varying wavelength coverages, spectral resolutions, and
signal-to-noise regimes of different surveys.

One widely-used strategy for mitigating inter-survey systematics is
\emph{label transfer}.
The idea is to train a model to predict the labels of a high-quality
``reference'' survey (typically APOGEE or GALAH) using spectra from a
larger but lower-resolution ``target'' survey, using stars observed
in both as a training set.
Once trained, the model is applied to all targets in the target
survey, delivering labels at the quality of the reference survey and
the scale of the target survey.
This approach has been applied across a range of survey combinations:
\citet{Ho2017label,Ho2017mass} transferred APOGEE labels to LAMOST
spectra using \textit{The Cannon} \citep{Ness2015};
\citet{Wheeler2020} transferred GALAH labels to LAMOST spectra;
\citet{Nandakumar2022} transferred APOGEE labels to GALAH spectra,
GALAH labels to APOGEE spectra;
\citet{Guiglion2024} transferred APOGEE labels to \emph{Gaia} RVS spectra;
and \citet{Das2025} transferred GALAH labels to \emph{Gaia} RVS spectra.
At a larger scale, the \emph{Survey of Surveys}
\citep{Tsantaki2022sos, Turchi2025sos} homogenizes stellar parameters
and radial velocities for ${\sim}$23 million stars from a variety of surveys.
However, this is done by post-hoc recalibration of existing pipeline
outputs rather than direct analysis of the spectra.

While these efforts have substantially extended the reach of
high-quality label sets, each study has performed \emph{pairwise}
transfer from one reference survey to one target survey.
To date, no work has delivered direct, simultaneous, homogeneous
stellar labels from spectra spanning more than two surveys at a time.

Here we present a unified deep-learning framework that addresses this
limitation.
Building on the spectral tokenizer architecture of
\citet{Shen2025tokenizer}---a 1D Vision Transformer capable of
encoding spectra of arbitrary wavelength range and resolution into a
common latent representation---we train a single model end-to-end to
predict atmospheric parameters, chemical abundances, distances, and
ages for all surveys simultaneously.
Because a single model processes all surveys jointly, every predicted
label is placed on a fully consistent scale by construction, without
requiring any post-hoc recalibration.

We demonstrate this framework by applying it to spectra from
\emph{four major surveys}: APOGEE DR17, GALAH DR3, DESI DR1, and
\emph{Gaia} RVS DR3.
Together these surveys span an enormous range of wavelength coverage
(near-infrared to optical), spectral resolution ($R \sim
1{,}800$--$28{,}000$), and stellar populations (metal-poor halo stars
to metal-rich disk dwarfs).
The resulting catalog delivers homogeneous labels for millions of
stars across the Milky Way, enabling Galactic archaeology at a scale
and consistency previously inaccessible.
This paper focuses on the framework, its validation, and the catalog
itself; a companion paper (Shen et al., in prep.) leverages this
catalog for detailed chemical cartography of the Milky Way.

This paper is structured as follows.
Section~\ref{sec:data} describes the spectroscopic surveys and
reference labels used in this work, along with our data preparation
and quality cuts.
Section~\ref{sec:method} presents the model architecture, training
procedure, and application to the full survey data.
Section~\ref{sec:validation} validates our results against held-out
data and comparison with existing pipeline labels and open cluster benchmarks.

\section{Data}
\label{sec:data}

\subsection{Spectroscopic surveys}

We use spectra from four large-scale surveys spanning wavelength
coverage from the near-infrared to the optical, resolutions from $R
\sim 1{,}800$ to ${\sim}\,28{,}000$, and a diverse range of stellar
populations and Galactic environments.
Key properties of each survey are summarized in Table~\ref{tab:surveys}.

\textbf{APOGEE DR17.}
The Apache Point Observatory Galactic Evolution Experiment
\citep[APOGEE;][]{Majewski2017} is an SDSS multi-object spectroscopic
survey that collected high-resolution ($R \sim 22{,}500$) spectra in
the near-infrared (1.51--1.70~$\mu$m), primarily targeting red giant
stars in high-extinction regions of the disc and bulge.

\textbf{GALAH DR3.}
The Galactic Archaeology with HERMES survey
\citep[GALAH;][]{buder_galah_2021} uses the HERMES spectrograph on
the 3.9~m Anglo-Australian Telescope.
HERMES simultaneously covers four optical wavelength windows
(4713--4903, 5648--5873, 6478--6737, and 7585--7887~\AA) at $R \sim
28{,}000$, designed to deliver detailed chemical abundances for up to
30 elements per star.
GALAH DR3 provides spectra and labels for ${\sim}600{,}000$ nearby
stars (both dwarfs and giants), most of which are within
${\sim}$4~kpc of the Sun.

\textbf{DESI DR1.}
The Dark Energy Spectroscopic Instrument \citep[DESI;][]{DESIDR1}
operates on the 4~m Mayall telescope at Kitt Peak National
Observatory, equipped with 5{,}000 robotically positioned fibers
feeding ten three-arm spectrographs covering 3600--9824~\AA.
The resolution varies from $R \sim 2{,}000$ in the blue arm to $R
\sim 5{,}500$ in the near-infrared arm.
While designed primarily as a redshift survey, DESI's Milky Way
Survey \citep{Cooper2023mws} also targets millions of stars,
preferentially selecting the thick disk and stellar halo.
We use the first public data release, DR1.

\textbf{\textit{Gaia} RVS DR3.}
The Radial Velocity Spectrometer \citep[RVS;][]{GaiaRVS} aboard the
\emph{Gaia} spacecraft \citep{GaiaDR3} is a slitless spectrograph
covering 846--870~nm (targeting the calcium triplet) at $R \sim 11{,}500$.
RVS targets bright stars ($G \lesssim 16$) in the \emph{Gaia} volume,
with a million mean RVS spectra released in DR3.

\begin{table}[ht]
  \centering
  \caption{Key properties of the spectroscopic surveys used in this work.}
  \label{tab:surveys}
  \begin{tabular}{llr}
    \toprule
    Survey & Wavelength range & $R$ \\
    \midrule
    APOGEE DR17         & 1.51--1.70~$\mu$m                    &
    22{,}500         \\
    GALAH DR3           & 4718--7890~\AA\ (4 windows)          &
    28{,}000         \\
    DESI DR1            & 3600--9824~\AA                       &
    2{,}000--5{,}500 \\
    Gaia RVS DR3        & 8460--8700~\AA                       &
    11{,}500         \\
    \bottomrule
  \end{tabular}
\end{table}

\subsection{Reference labels}

As ground-truth labels for training our model, we use the
\texttt{astroNN} value-added catalog (VAC) for APOGEE
DR17,\footnote{\url{https://www.sdss4.org/dr17/data_access/value-added-catalogs/?vac_id=the-astronn-catalog-of-abundances,-distances,-and-ages-for-apogee-dr17-stars}}
which provides atmospheric parameters, 20 individual elemental
abundances, distances, and stellar ages for APOGEE red giants
\citep{leung_deep_2018, mackereth_dynamical_2019}.
Labels were derived using the \texttt{astroNN} deep learning
framework \citep{leung_deep_2018}, trained on atmospheric parameters
and abundances from the APOGEE Stellar Parameters and Chemical
Abundances Pipeline (ASPCAP) DR17.
Stellar ages were trained on precise asteroseismic ages from the
APOKASC-2 catalog \citep{Pinsonneault2018}, which cross-matches
APOGEE targets with red giants observed by the \emph{Kepler} space
telescope \citep{mackereth_dynamical_2019}.
Since our model learns to predict these labels from spectra across
all surveys simultaneously, the resulting labels for every survey are
automatically placed on this common scale by construction, without
any post-hoc recalibration.

\subsection{Data preparation}

\textbf{Spectra quality cuts.}
For APOGEE we require a per-spectrum signal-to-noise ratio \texttt{snr > 10}.
For GALAH we require \texttt{snr\_c3\_iraf > 30} following
recommendations from the GALAH team.
For DESI we require stellar classification (\texttt{SPECTYPE =
`STAR'}), no pipeline warnings (\texttt{ZWARN = 0}), a reliable
redshift (\texttt{ZERR} ${<}\,10^{-4}$), a template fit quality
\texttt{DELTACHI2 > 100}, and a detection with signal-to-noise
${>}\,5$ in at least one of the three spectrograph arms
(\texttt{MEDIAN\_COADD\_SNR\_B}, \texttt{MEDIAN\_COADD\_SNR\_R}, or
\texttt{MEDIAN\_COADD\_SNR\_Z}).
No quality cuts are applied to the \emph{Gaia} RVS spectra.

\textbf{Label quality cuts.}
We additionally require well-constrained AstroNN label estimates.
For atmospheric parameters we require effective temperature
uncertainty of less than $10\%$, surface gravity uncertainty
$\texttt{LOGG\_ERR} < 0.3$~dex, and iron abundance uncertainty
$\texttt{FE\_H\_ERR} < 0.2$~dex.
For both distances and ages we require a valid, positive estimate,
and fractional uncertainty less than $30\%$.

We cross-match the filtered spectra from each survey against AstroNN
VAC using a 1-arcsec radius search.
The resulting cross-matched dataset is split 90/10 into training and
validation subsets; final sizes for each survey are listed in
Table~\ref{tab:dataset_comparison}.

\begin{table}[h]
  \centering
  \caption{Training and validation set sizes by dataset after quality
  cuts and cross-matching.}
  \label{tab:dataset_comparison}
  \begin{tabular}{lrr}
    \hline
    \textbf{Dataset} & \textbf{Train Size} & \textbf{Val Size} \\ \hline
    APOGEE DR17                          & 199,382            &
    22,153           \\
    GALAH DR3                            & 12,125             & 1,347
    \\
    DESI DR1                             & 5,952              & 661
    \\
    Gaia RVS DR3                         & 80,600             & 8,955
    \\ \hline
    \textbf{Total}                       & \textbf{298,289}   &
    \textbf{33,141}  \\ \hline
  \end{tabular}
\end{table}

\section{Method}
\label{sec:method}

We build on the spectral tokenizer architecture of
\citet{Shen2025tokenizer}, which uses a 1D Vision Transformer-style
autoencoder \citep{Dosovitskiy2021vit} to ingest spectrum flux,
uncertainty, and wavelength, compress the input into a sequence of
latent ``tokens,'' and reconstruct the original spectrum.
A key property of this architecture is its ability to ingest spectra
of arbitrary resolution and wavelength range, enabling a single model
to process heterogeneous data from multiple surveys simultaneously
and to transform it into a homogeneous representation.
Here we use only the encoder component and train it end-to-end with a
lightweight regression head, rather than following the two-stage
approach (self-supervised pre-training of encoder and decoder,
followed by frozen encoder and a task-specific head) described in
\citet{Shen2025tokenizer}.
We adopt this strategy because the existing pre-trained model does
not cover all the surveys used in the present work; we plan to
release a model pre-trained on a much larger set of spectroscopic
surveys in the future.

\newcommand\rotnode[5]{%
  \node [#1, opacity=1.0] (#3) {#4};
  \node [#2, rotate around={#4:(#3.center)}] at (#3) (#3_box) {#5};
}

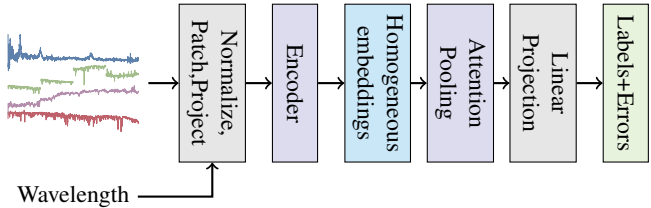
\begin{figure}[t]
  \centering
  \begin{tikzpicture}[
      layer/.style={rectangle, draw=black, fill=#1!20, minimum
      width=2.1cm, minimum height=0.6cm, align=center, font=\small},
      arrow/.style={->, thick},
      every node/.style={font=\small}
    ]

    \rotnode{}{layer=gray}{patch}{270}{Normalize,\\Patch,Project};
    \rotnode{right=0.4cm of patch}{layer=Periwinkle}{encoder}{270}{Encoder};
    \rotnode{right=0.4cm of
    encoder}{layer=Cerulean}{embeddings}{270}{Homogeneous\\embeddings};
    \rotnode{right=0.4cm of
    embeddings}{layer=Periwinkle}{pooling}{270}{Attention\\Pooling};
    \rotnode{right=0.4cm of
    pooling}{layer=gray}{project}{270}{Linear\\Projection};
    \rotnode{right=0.4cm of
    project}{layer=YellowGreen}{labels}{270}{Labels+Errors};

    \node[inner sep=0pt, left=0.5cm of patch] (spec_in)
    {\includegraphics[width=.11\textwidth]{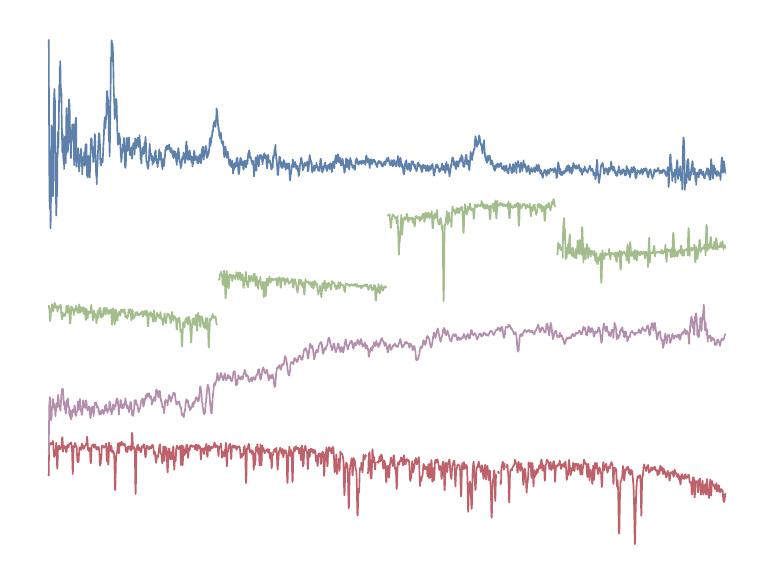}};

    \node[below=0.5cm of spec_in] (wavelength) {Wavelength};

    \draw[arrow] (spec_in) -- (patch_box);
    \draw[arrow] (patch_box) -- (encoder_box);
    \draw[arrow] (encoder_box) -- (embeddings_box);
    \draw[arrow] (embeddings_box) -- (pooling_box);
    \draw[arrow] (pooling_box) -- (project_box);
    \draw[arrow] (project_box) -- (labels_box);

    \coordinate (emb_anchor) at ([yshift=0.25cm]embeddings_box.west);

    \draw[arrow] (wavelength) -| (patch_box.east);

  \end{tikzpicture}
  \caption{Brief overview of the model architecture. The encoder
    creates homogeneous, wavelength-aware embeddings from heterogeneous
    input spectra; these embeddings are then pooled and mapped to
  stellar labels. Figure adapted from \citet{Shen2025tokenizer}.}
  \label{fig:model_architecture}
\end{figure}

The architecture of our model is shown in Figure~\ref{fig:model_architecture}.
We refer the reader to \citet{Shen2025tokenizer} for more a detailed
technical description of the encoder architecture; here we provide a
brief overview and primarily describe the regression head.
Input spectra are first normalized by their median, then stacked with
the inverse variance weights, chunked into patches of $P$ contiguous
pixels, and linearly projected into a sequence of embedding vectors
following the standard procedure for Vision Transformers
\citep{Dosovitskiy2021vit}.
The median is preserved as a separate vector, which is prepended to
the sequence of embeddings.
A sinusoidal encoding of the wavelength axis is similarly patched and
elementwise added to each patch embedding to imbue the embeddings
with wavelength position information.
The resulting sequence is processed by a stack of Transformer encoder
blocks with full quadratic self-attention
\citep{vaswani_attention_2017}, SwiGLU activations
\citep{shazeer2020}, QK-normalization \citep{henry2020qknorm}, and
Rotary Position Embeddings (RoPE;
\citealt{su2023roformerenhancedtransformerrotary}).
A cross-attention pooling layer with a learnable query vector
aggregates the token sequence into a single vector.
Notably, it is at this stage that all spectra, regardless of their
resolution and wavelength range at the input level, are fully
homogenized: they are all represented as a single vector of the same
dimensionality.
A final linear layer projects this vector, or ``embedding,'' to the label space.

For each label $i$ the model produces two outputs: a predicted value
$\hat{y}_i$ and a predicted uncertainty $\hat{\sigma}_i$.
We train with the heteroscedastic Gaussian negative log-likelihood
loss function, also used in, for example, \citet{leung_deep_2018} and
\citet{Leung_2023}:
\begin{align}
  \mathcal{L} = \frac{1}{2} \left(\frac{(y_i -
    \hat{y}_i)^2}{\sigma^2_i + \hat{\sigma}^2_i} + \ln{(\sigma^2_i +
  \hat{\sigma}^2_i)} \right),
\end{align}
where $y_i$ and $\sigma_i$ are the ground-truth label value and its
associated measurement uncertainty.
For numerical stability we parameterize $\hat{\sigma}_i$ in
log-space, so that the network outputs $\ln(\hat{\sigma}_i)$ and we
exponentiate where needed.

We additionally train with dropout enabled in both the attention and
feedforward layers, and apply Monte Carlo Dropout \citep{Gal2016} at
inference time: $k$ forward passes are performed with dropout
enabled, yielding an ensemble of predictions $\{\hat{y}_{i,j},
\hat{\sigma}_{i,j}\}_{j=1}^{k}$.
The total predictive variance is the sum of the mean aleatoric
variance (predicted by the model) and the epistemic variance (due to
model uncertainty):
\begin{align}
  \label{eq:model-uncertainty}
  \sigma_{\mathrm{total}, i}^2 = \frac{1}{k} \sum_{j=1}^k
  \hat{\sigma}_{i,j}^2 + \frac{1}{k} \sum_{j=1}^k (\hat{y}_{i,j} -
  \bar{\hat{y}}_i)^2,
\end{align}
where $\bar{\hat{y}}_i = \frac{1}{k}\sum_{j=1}^{k}\hat{y}_{i,j}$ is
the mean of the predicted means across the ensemble.

We train a \textit{single model} jointly on all four surveys and all
output labels.
At each training step a survey is selected randomly (with the
  selection probability proportional to the number of samples in the
survey) and a full batch is drawn from that survey; this ensures that
all spectra within a batch share the same wavelength grid, avoiding
the need for padding.
We use an embedding dimension of 384, feed-forward hidden dimension
of 1536, 6 Transformer blocks, 8 attention heads, and a patch size of 32 pixels.
The model has a total parameter count of ${\sim}15$~M.
We train on four NVIDIA H100 GPUs for 500{,}000 steps using AdamW
($\beta_1 = 0.9$, $\beta_2 = 0.95$) with a constant learning rate of
$2 \times 10^{-4}$, no weight decay, a dropout rate of 0.1, and
$k=32$ MC dropout forward passes.

\subsection{Application}
\label{subsec:application}

After training, we apply the model to all spectra from the four
surveys that pass stricter spectrum SNR quality cuts than the ones
described in Section~\ref{sec:data}.
For APOGEE we require \texttt{snr > 75} and that \texttt{STAR\_BAD =
0} in the \texttt{ASPCAPFLAG} bitmask.
For \textit{Gaia} RVS we require \texttt{rvs\_spec\_sig\_to\_noise > 30}.
For DESI we require that at least one of
\texttt{MEDIAN\_COADD\_SNR\_\{B,R,Z\} > 30}, in addition to the other
cuts described in Section~\ref{sec:data}.
We additionally apply our model to SDSS DR17 optical spectra
(Legacy/SEGUE/BOSS/eBOSS; \citealt{York2000, Yanny2009, Dawson2013,
Dawson2016}) that are spectroscopically classified as stars
(\texttt{CLASS = STAR}), have no pipeline warnings (\texttt{ZWARN =
0}), have a reliable redshift (\texttt{ZERR} ${<}\,10^{-4}$), have
good plate quality (\texttt{PLATEQUALITY = good}), and a median SNR
of \texttt{SN\_MEDIAN\_ALL} ${>}\,30$.
The resulting catalog provides homogeneous stellar labels for every
star in the combined sample, with sample counts provided in
Table~\ref{tab:application_counts}.

\begin{table}[h]
  \centering
  \caption{Number of spectra from each survey that pass the quality
    cuts described in Section~\ref{subsec:application} and are included
  in the final catalog.}
  \label{tab:application_counts}
  \begin{tabular}{lr}
    \hline

    \textbf{Dataset} & \textbf{Count} \\ \hline
    APOGEE DR17                          & 454,151         \\
    DESI DR1                            & 1,827,690          \\
    GALAH DR3                             & 419,318           \\
    \textit{Gaia} RVS DR3                         & 469,618          \\
    SDSS DR17 (Legacy/SEGUE/BOSS/eBOSS) & 175,079        \\ \hline
    \textbf{Total}                       & \textbf{3,345,856}       \\ \hline
  \end{tabular}
\end{table}

\subsection{Flagging}
\label{subsec:flagging}

We provide several ways to determine whether predicted quantities are
reliable, both on the input and output side.
On the output (label) side, the predicted model uncertainty is a good
indication of when the model prediction is reliable.
We additionally flag values where the predicted label lies below the
1st percentile or above the 99th percentile of the distribution of
that label in the training set; performance tends to suffer from edge
effects (i.e., clamping) near the limits of the label distribution.

On the input/spectrum side, we provide three indicators of reliability.
These are all measures of how out-of-distribtion (OOD) the spectrum
that the model is being applied to is, relative to the spectra that
were used to train the model.
They are all based on the embeddings of the spectrum, as calculated
by the model.
Here, embeddings are calculated by taking the vector representation
of the spectrum after it has passed through all layers of the model
except for the final projection into label space.
This is done for the entire training set, with dropout disabled.

The first metric, and the simplest, is the Mahalanobis distance of
the embeddings to the training distribution of embeddings \citep{lee2018ood}.
This is calculated by fitting a full covariance matrix to the
distribution of training embeddings.
At inference time, we use this covariance matrix to determine the
Mahalanobis distance of any particular test spectrum, and report this
as an additional column.
We warn, however, that the embedding manifold is not structured in a
way that makes Mahalanobis distance the best choice as an OOD
detection method (i.e., Gaussian).
Furthermore, distances tend to concentrate in high dimensions,
leading to a reduction in contrast between in-distribution and
out-of-distribution samples \citep[e.g.,][]{Aggarwal2001}.
As such, the Mahalanobis distance should be used with caution as an
OOD detection method, and we recommend using it in conjunction with
the other metrics, described below.
The second metric is a score from an Isolation Forest model, a
nonparametric OOD detector \citep{Liu2008isolationforest}, which we
also report as an additional column.
The third metric is a likelihood from a generative model, namely a
Gaussian Mixture Model (GMM), fit to the training embeddings with
$K=10$ components to ensure flexibility in capturing the structure of
the embedding manifold.
Generative models have been used for OOD/anomaly detection in other
works within astronomy, and in particular by
\citet{Liang2023outlier}, who used normalizing flow likelihoods on
embeddings of galaxy spectra for outlier detection.
For all of these metrics, we also provide calibrated measures: for
the Mahalanobis distance/Isolation Forest score/GMM likelihood, we
calculate the percentile at which the test spectrum's
distance/score/likelihood is at, relative to the distribution of
distances/scores/likelihoods in the reference set.
This allows for flexible and easy downstream filtering based on how
conservative one would like to be with OOD detection by making cuts
at different percentiles.

\section{Validation}
\label{sec:validation}

We quantify performance using $\sigma_{\rm MAD} = 1.4826~{\rm MAD}$,
where ${\rm MAD} = \texttt{median}(|e_i - \texttt{median}(e)|)$ is a
robust measure of standard deviation for a set of residuals $e =
\{e_1, \ldots, e_n\}$, and the 1.4826 factor is so that for a
Gaussian distribution, $\sigma_{\rm MAD} = \sigma$, the usual measure
of standard deviation.

\subsection{Internal validation}

\begin{figure*}[t]
  \centering
  \includegraphics[width=0.9\linewidth]{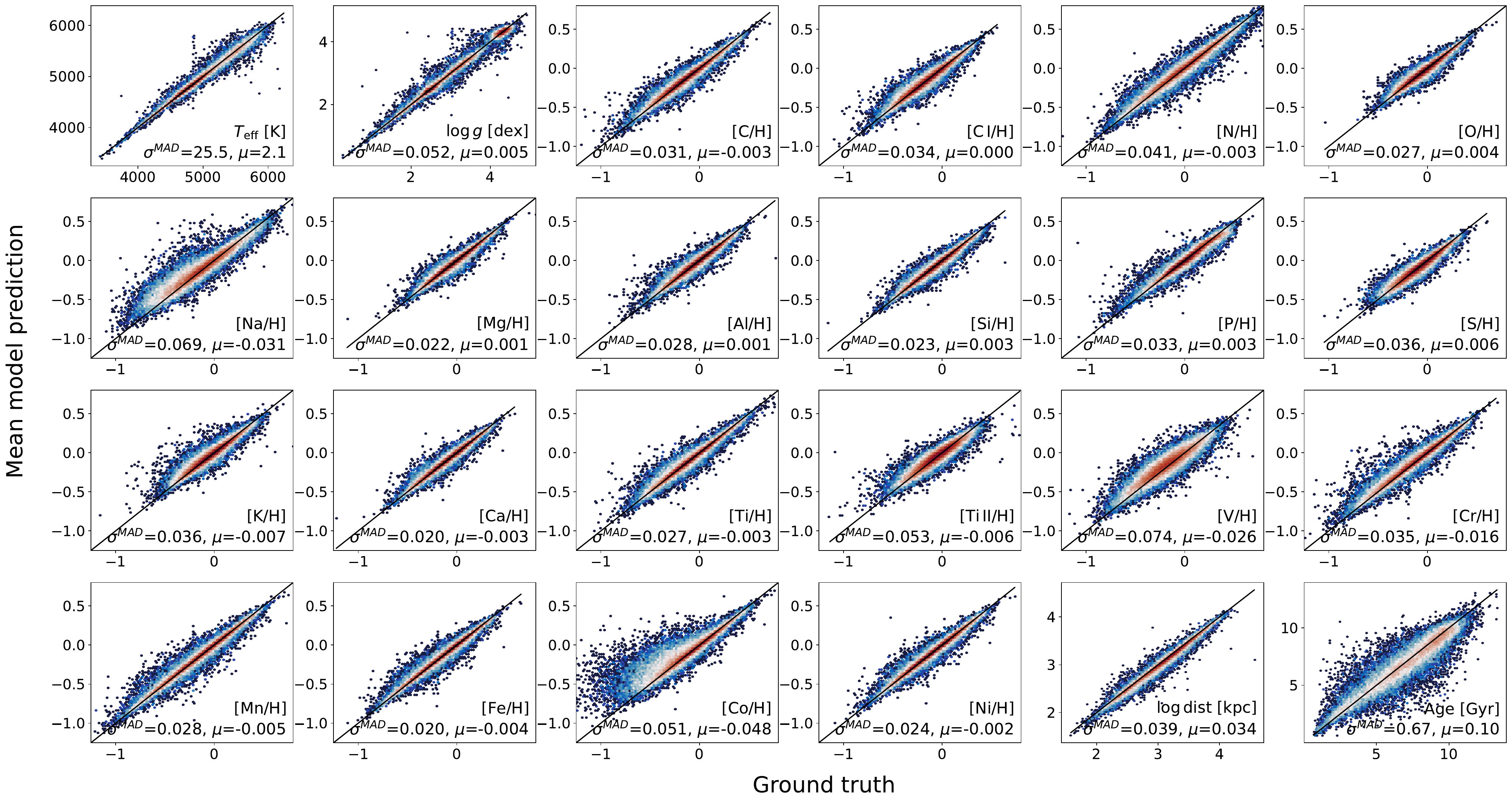}
  \caption{Ground truth label values (x-axes) vs predicted label
    values (y-axes) for atmospheric parameters, chemical abundances,
    distances, and stellar ages, aggregated across all samples in the
    validation dataset and coloured by log density (red indicates more
    samples in that region, and blue indicates fewer samples). Black
    lines indicate 1-to-1 relations; in general the model predictions
  are very tightly clustered around these lines.}
  \label{fig:eval_all}
\end{figure*}

We perform internal validation of the model by evaluating it on the
held-out validation set.
Figure \ref{fig:eval_all} shows the mean model prediction (across the
MC Dropout ensemble) plotted against the ground truth labels from AstroNN.
We find that the model accurately and precisely estimates atmospheric
parameters, distances, and ages, and nearly all abundances.
The points are clustered tightly around the 1-to-1 line, indicating
that most labels are well estimated.
Two exceptions are [Co/H] and [Na/H], where our model exhibits
systematic overestimation, particularly at the metal-poor end;
we note that \citep{leung_deep_2018} also find difficulty in
predicting [Co/H], [P/H], [Na/H], and [TiII/H], and their model
produces larger uncertainties for those elements.

We decompose the per-label model performance into the constituent
surveys in Table \ref{tab:performance-by-survey}.
We find that the model performs extremely well on APOGEE spectra;
the typical precision on key atmospheric parameters is $18$~K on
$\teff$, $0.04$~dex on $\logg$, and $0.015$~dex on $\feh$.
Across all chemical abundances that are estimated by the model, the
mean precision is ${<}\,0.03$~dex.
On larger datasets of lower spectral resolution, like DESI, the model
still is extremely capable: typical precision is $51$~K on $\teff$,
$0.09$~dex on $\logg$, and $0.041$~dex on $\feh$.
Mean precision on chemical abundances is ${\sim}\,0.06$~dex.
\begin{table}[!htbp]
  \centering
  \begin{tabular}{lcccc}
    \toprule
    \textbf{Label} & \textbf{\textit{Gaia} RVS} & \textbf{APOGEE} &
    \textbf{GALAH} & \textbf{DESI} \\
    \midrule
    $T_{\mathrm{eff}}$ [K] & 56.2  & 18.2  & 52.7  & 51.0  \\
    $\log g$ [dex]         & 0.102 & 0.040 & 0.101 & 0.092 \\
    $[\mathrm{C/H}]$       & 0.058 & 0.025 & 0.050 & 0.050 \\
    $[\mathrm{C\,I/H}]$    & 0.061 & 0.027 & 0.058 & 0.053 \\
    $[\mathrm{N/H}]$       & 0.073 & 0.033 & 0.069 & 0.063 \\
    $[\mathrm{O/H}]$       & 0.048 & 0.022 & 0.043 & 0.048 \\
    $[\mathrm{Na/H}]$      & 0.085 & 0.062 & 0.086 & 0.101 \\
    $[\mathrm{Mg/H}]$      & 0.047 & 0.017 & 0.038 & 0.037 \\
    $[\mathrm{Al/H}]$      & 0.050 & 0.022 & 0.044 & 0.046 \\
    $[\mathrm{Si/H}]$      & 0.048 & 0.017 & 0.039 & 0.044 \\
    $[\mathrm{P/H}]$       & 0.053 & 0.027 & 0.049 & 0.054 \\
    $[\mathrm{S/H}]$       & 0.051 & 0.031 & 0.052 & 0.058 \\
    $[\mathrm{K/H}]$       & 0.057 & 0.029 & 0.051 & 0.055 \\
    $[\mathrm{Ca/H}]$      & 0.044 & 0.016 & 0.034 & 0.036 \\
    $[\mathrm{Ti/H}]$      & 0.059 & 0.021 & 0.045 & 0.048 \\
    $[\mathrm{Ti\,II/H}]$  & 0.068 & 0.047 & 0.071 & 0.062 \\
    $[\mathrm{V/H}]$       & 0.090 & 0.067 & 0.095 & 0.105 \\
    $[\mathrm{Cr/H}]$      & 0.062 & 0.028 & 0.051 & 0.056 \\
    $[\mathrm{Mn/H}]$      & 0.062 & 0.021 & 0.048 & 0.055 \\
    $[\mathrm{Fe/H}]$      & 0.049 & 0.015 & 0.037 & 0.041 \\
    $[\mathrm{Co/H}]$      & 0.082 & 0.039 & 0.086 & 0.129 \\
    $[\mathrm{Ni/H}]$      & 0.050 & 0.019 & 0.039 & 0.042 \\
    $\log{\rm dist}$ [kpc] & 0.067 & 0.028 & 0.124 & 0.038 \\
    Age [Gyr]              & 1.081 & 0.554 & 0.983 & 0.819 \\
    \bottomrule
  \end{tabular}
  \caption{Model precision ($\sigma_{\rm MAD}$ of residuals against
  ground truth labels) for each label, decomposed by survey.}%
  \label{tab:performance-by-survey}
\end{table}

\begin{figure*}[t]
  \centering
  \includegraphics[width=1\linewidth]{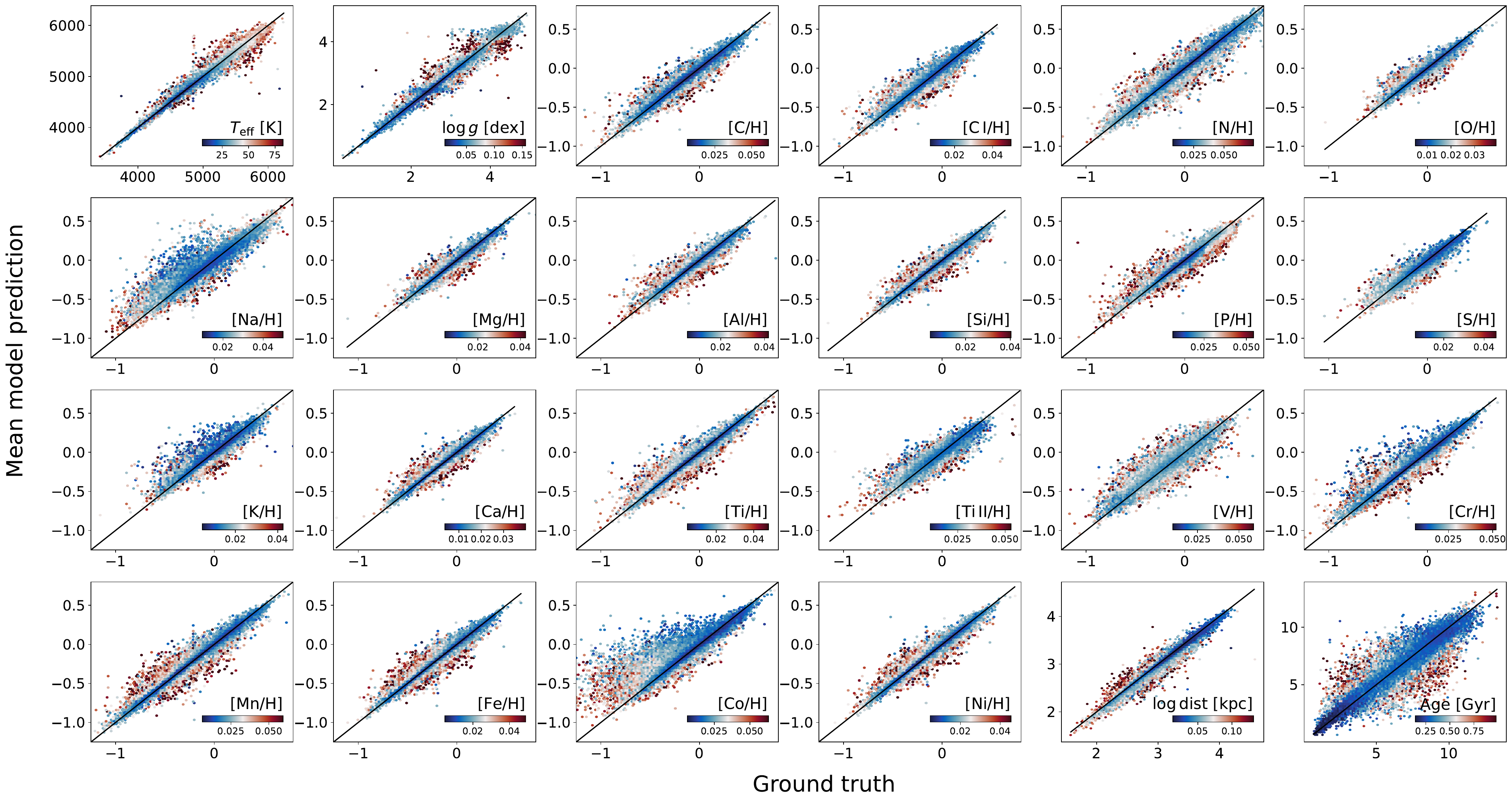}
  \caption{Same as Figure \ref{fig:eval_all}, but coloured by the
    predicted uncertainty from the model, as indicated by the inset
    colorbars in each panel. The model predicts higher uncertainty in
  regions further from the 1-to-1 line.}
  \label{fig:eval_all_by_err}
\end{figure*}

Figure \ref{fig:eval_all_by_err} is a similar plot to Figure
\ref{fig:eval_all}, but the colour indicates the total model
uncertainty (see Equation \ref{eq:model-uncertainty}) rather than
sample density.
Assuringly, points near the 1-to-1 line, which are accurately
predicted, are also associated with low uncertainties, while points
further from the diagonal are associated with larger uncertainties.
This is an indication that the model has a good understanding of
where it is uncertain.

\begin{figure*}[t]
  \centering
  \includegraphics[width=1\linewidth]{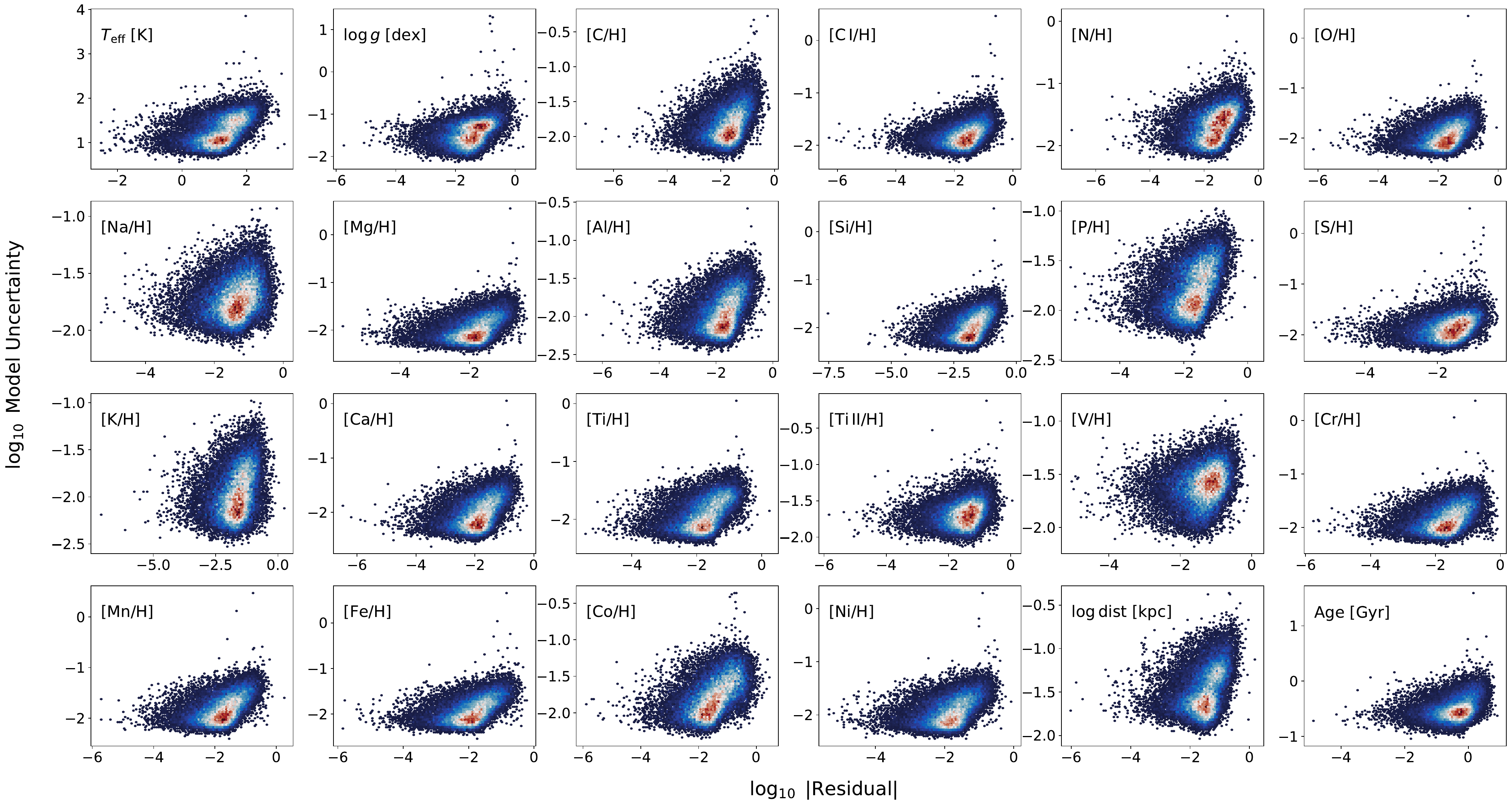}
  \caption{Residuals (x-axis) vs total model uncertainty, aggregated
    across all samples in the validation dataset and coloured by log
    density (red indicates more samples in that region, and blue
    indicates fewer samples). Across all labels, the predictive
  uncertainty from the model is positively correlated with actual residuals.}
  \label{fig:residual_vs_error}
\end{figure*}

We show in Figure \ref{fig:residual_vs_error} the residuals plotted
against the predicted model error; in general these two quantities
are positive correlated, further indicating that the model
uncertainties are a reasonable proxy for actual estimation error.
We perform a more detailed characterization of the uncertainties in
Section \ref{subsec:uncertainty_calibration}.

\begin{figure}[ht]
  \centering
  \includegraphics[width=1\linewidth]{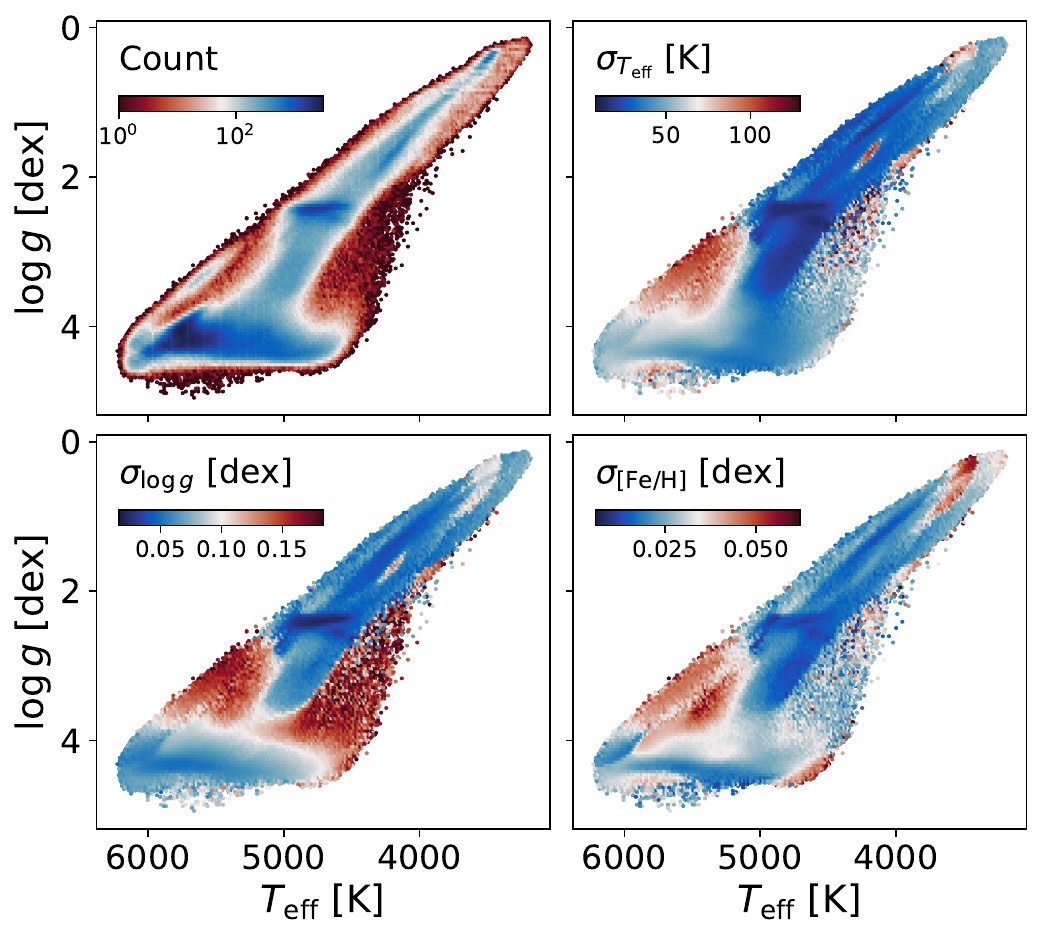}
  \caption{Model uncertainty across parameter space. The top-left
    panel shows a Kiel diagram coloured by sample density (red
    indicates fewer samples), while the remaining panels show the same
    diagram coloured by the predicted uncertainty in $\teff$ (top
    right), $\logg$ (bottom left), and $\feh$ (bottom right). Note that
    the colour scale for sample density is inverted relative to the
  uncertainty panels to facilitate comparison.}
  \label{fig:kiel_by_error}
\end{figure}

In Figure \ref{fig:kiel_by_error} we show model uncertainty across
parameter space.
The top left Kiel diagram is coloured by sample density, with red
regions indicating few samples, and the other plots are coloured by
$\teff$ uncertainty (top right), $\logg$ uncertainty (bottom left),
and $\feh$ uncertainty (bottom right).
Note that we have inverted the colours for the top-left panel (sample
density); this is to make comparison across the panels easier, since
model uncertainty correlates strongly with the amount of training
data available in a given region of parameter space.
We find that uncertainties are larger for cooler stars, and
particularly so for cool dwarfs.
On the other hand, parameters appear to be generally well constrained
for giants, and red clump stars appear as an obvious regime in
parameter space for which uncertainties are very low.
This makes sense given that our reference label scale is derived from
APOGEE, which mainly targets red giants.

\subsection{Uncertainty calibration}
\label{subsec:uncertainty_calibration}

While the model's predicted uncertainties are positively correlated
with actual residuals (Section~\ref{sec:validation}), this alone does
not guarantee that the uncertainties are well
\emph{calibrated}---that is, that a predicted $1\sigma$ interval
contains the true value ${\sim}\,68\%$ of the time.
To assess calibration, we use the probability integral transform
\citep[PIT;][]{Dawid1984, Gneiting2007}.
For each validation-set prediction $i$, we compute the PIT value
\begin{align}
  p_i = \Phi\!\left(\frac{y_i -
    \hat{y}_i}{\sqrt{\sigma_{\mathrm{pred},i}^2 +
  \sigma_{\mathrm{label},i}^2}}\right),
\end{align}
where $\Phi$ is the cumulative distribution function of the standard
normal, $\sigma_{\mathrm{pred},i}$ is the total predictive
uncertainty from the model, and $\sigma_{\mathrm{label},i}$ is the
uncertainty on the training label for that sample.
Intuitively, $p_i$ is the quantile of the predicted distribution at
which the true value falls: if a star's true label sits at the median
of the predictive distribution then $p_i = 0.5$, while values near
$0$ or $1$ indicate the true label lies in the tails.
If the predictive distributions are perfectly calibrated, the PIT
values are uniformly distributed on $[0, 1]$, regardless of the shape
of the underlying predictive distribution.
Deviations from uniformity reveal specific miscalibration patterns: a
U-shaped PIT histogram indicates underdispersed (overconfident)
uncertainties, an inverted-U shape indicates overdispersed
(conservative) uncertainties, and skewness indicates a systematic
bias in the predictions.
This has been used in other works in astrophysics to assess the
calibration of predictive posteriors for regression tasks
\citep[e.g.,][]{Shen2024disentangling}.

\begin{figure*}[ht]
  \centering
  \includegraphics[width=1\linewidth]{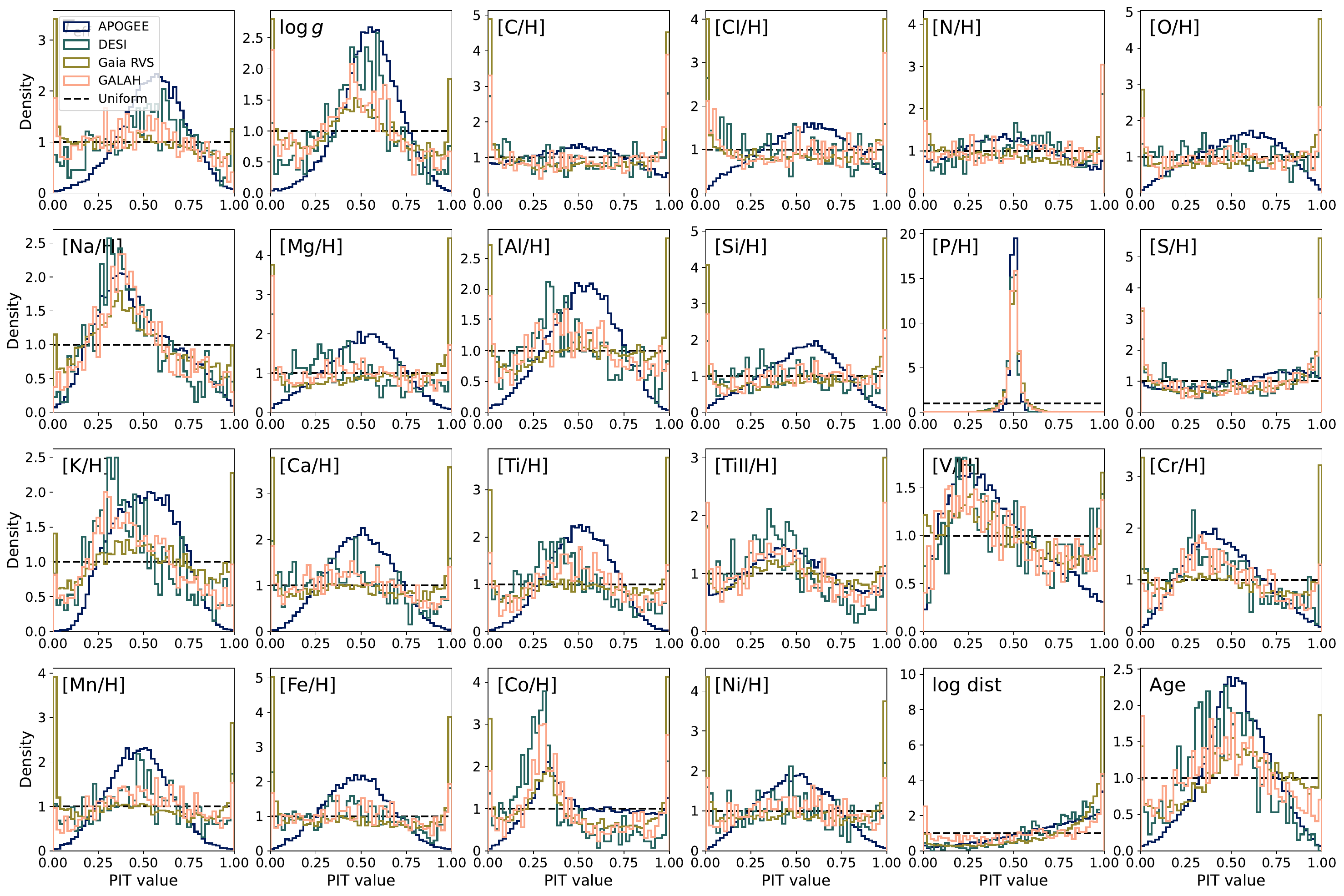}
  \caption{PIT histograms for all labels, broken down by survey. A
    perfectly calibrated model would produce a uniform distribution
    (dashed black line). A U-shaped histogram indicates underestimated
    uncertainties, while an inverted-U shape indicates overestimated
  uncertainties.}
  \label{fig:calibration}
\end{figure*}

Figure~\ref{fig:calibration} shows PIT histograms for estimated
labels, decomposed by survey.
We find that the calibration quality varies systematically across
surveys and labels, with several clear patterns.

For APOGEE, the majority of labels exhibit inverted-U shaped PIT
histograms, indicating that the model uncertainties are overly conservative.
The predicted uncertainty intervals are wider than necessary, so that
the true values cluster toward the center of the predicted
distribution more often than a well-calibrated model would produce.
We note that this is actually due to the fact that the underlying
AstroNN label uncertainties are themselves overly conservative; we
point to Figure \ref{fig:calibration_labelerr_only}, which shows PIT
histograms calculated using only the label uncertainties (i.e.,
without the model predictive uncertainty), where the same inverted-U
shapes are present for APOGEE.
Given that the total model uncertainty is the sum in quadrature of
the model predictive uncertainty (from our model) and the label error
(from AstroNN), we cannot expect to achieve less conservative
uncertainties than the label uncertainties alone.

\textit{Gaia} RVS is in general more well calibrated than APOGEE,
with many labels showing PIT histograms that are closer to uniform.
However, for almost all labels, the PIT histograms for \textit{Gaia}
RVS show spikes around $p=0$ and $p=1$, indicating that there are a
non-negligible number of catastrophic outliers for which the true
value lies far in the tails of the predicted distribution, and these
outliers are not captured by the model uncertainty.
That is, the uncertainty is massively underestimated for these
outliers, even if it is reasonably well calibrated for the bulk of
the distribution.
Examining Figure \ref{fig:calibration_labelerr_only}, we see again
that this behaviour is inherited from the label uncertainties, which
also show spikes near $p=0$ and $p=1$ for \textit{Gaia} RVS.
DESI and GALAH are closer to uniform than APOGEE, and do not exhibit
the massive spikes near $p=0$ and $p=1$ that are present for
\textit{Gaia} RVS, indicating that the model uncertainties are better
calibrated for these surveys than for \textit{Gaia} RVS.
However, the overestimation of uncertainties is still visible for
many labels (e.g., $\logg$, [Na/H], [V/H]).
In general, there do not appear to be labels for which DESI and GALAH
uncertainties are worse calibrated than for APOGEE.

Several individual labels are worth highlighting.
For [P/H], which is known to be highly unreliably in ASPCAP DR17, all
surveys shows an extreme spike near $p = 0.5$, indicating that the
predicted uncertainties are so large relative to the actual residuals
that essentially all true values fall near the center of the
predicted distribution.
[Co/H], unlike the other chemical abundances which peak around
$p=0.5$, indicating a lack of bias, has instead a peak around
$p=0.3$, indicating systematic overestimation for many values.
This is particularly pronounced for DESI, and this effect is also
visible in Figure \ref{fig:eval_all}, where predicted values of
[Co/H] show increasingly large scatter and overestimation towards
lower reference [Co/H] values from AstroNN.
For log distance, on the other hand, the PIT histogram is skewed
towards $p=1$, indicating a tendency towards systematic underestimation.
This is again visible in Figure \ref{fig:eval_all}, where a bias of
$0.039~$dex in the residuals of log distance is noted.

In summary, the PIT analysis reveals that the model uncertainties are
generally somewhat conservative, rather than overconfident; this
means that quoted error bars are unlikely to understate the true uncertainty.
Much of this conservatism is inherited from the label uncertainties,
which are themselves conservative, particularly for APOGEE.
However, there are some labels and surveys for which the model
uncertainties are not well calibrated, particularly for \textit{Gaia}
RVS, where there are a non-negligible number of catastrophic outliers
that are not captured by the model uncertainty, and for [P/H], where
the uncertainties are so large that they are essentially uninformative.
We recommend that users of the catalog be mindful of these
calibration issues when using the model uncertainties for downstream
analyses, and to consider applying additional cuts on the predicted
uncertainties for labels/surveys where the calibration is poor, if
they want to ensure a more conservative sample.

\begin{figure*}[ht]
  \centering
  \includegraphics[width=1\linewidth]{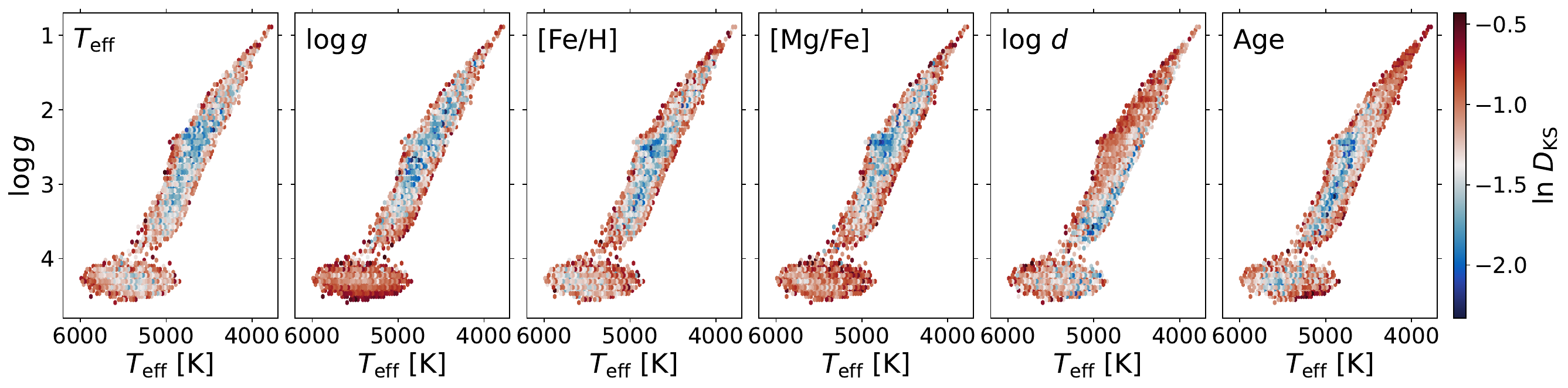}
  \caption{Calibration quality of predicted uncertainties across Kiel
    diagram for selected labels, aggregated across all input surveys.
    Colour indicates $\ln D_{\rm KS}$, the logarithm of the
    Kolmogorov–Smirnov statistic of the PIT distribution against a
    uniform reference distribution. More negative values (darker blue)
  indicate better calibrated uncertainties.}
  \label{fig:calibration_kiel}
\end{figure*}

We also show in Figure \ref{fig:calibration_kiel} an examination of
how well our predicted uncertainties are calibrated as a function of
stellar parameters.
Again, if the uncertainties are well calibrated and the residuals are
Gaussian, the PIT values follow a uniform distribution.
We thus calculate the Kolmogorov–Smirnov (KS) statistic of the PIT
distribution against a uniform reference distribution across the Kiel
diagram for selected labels, aggregated across all surveys.
We colour the Kiel diagram by the logarithm (for better dynamic
range) of the KS statistic, $\ln D_{\rm KS}$, where more negative
values (darker blue) indicate better calibration.
For [Mg/Fe] we propagate the [Mg/H] and [Fe/H] uncertainties in
quadrature, assuming independence.
PIT values are binned in hexagonal cells in $\teff-\logg$ space, and
we require at least five samples per bin.
We observe that the predicted uncertainties are generally better
calibrated for giants than for dwarfs, with particularly good
calibration for red clump stars, especially for [Fe/H] and [Mg/Fe].
This is unsurprising given that the reference label scale is derived
from APOGEE, which mainly targets red giants.
We also find that the calibration is generally correlated with the
amount of training data available in a given region of parameter
space (see the top-left panel of Figure \ref{fig:kiel_by_error}),
with worse calibration near the edges of the training distribution.
Interestingly, we find a metallicity-dependent pattern in the
calibration of the distances for giant stars: for metal-poor giants
(closer to the left edge of the Kiel diagram), the uncertainties are
more poorly calibrated than for metal-rich giants (closer to the
right edge of the Kiel diagram).

\subsection{Cross-survey validation}
\label{subsec:cross-survey}

As an additional sanity check, we compare values estimated for the
same object from two different surveys/spectra.
This is to check for self-consistency in the model; for the same
object, the model should predict the same output labels, even if the
input spectra are different.
We note that in theory, because our model is capable of ingesting
multiple input spectra, and the cross-attention pooling step is
agnostic to the input sequence length, we could potentially combine
multiple spectra into a single estimate where multiple spectra are
available for a given object, giving us stronger constraints on the labels.
We however do not do this in this work, opting instead for the
simpler option of predicting one set of labels for every spectrum,
rather than for every star.

We do the self-consistency check on the values from the output
catalog, because there is no need to limit the sample to the small
held-out validation set for this check.
We apply a few loose quality cuts, similar to that which a downstream
user might use for science applications.
In particular, we require that $\sigma_{\teff} < 150$~K,
$\sigma_{\logg} < 0.2$, $\sigma_{\feh} < 0.1$, and additionally that
the spectrum embedding is below the 99.5th percentile in Mahalanobis
distance, above the 2.5th percentile in Isolation Forest score, and
above the 0.5th percentile in GMM likelihood (see Section
\ref{subsec:flagging}).
We then cross-match objects that appear between pairs of surveys
using a 1-arcsecond radius search.

\begin{figure*}[t]
  \centering
  \includegraphics[width=0.9\linewidth]{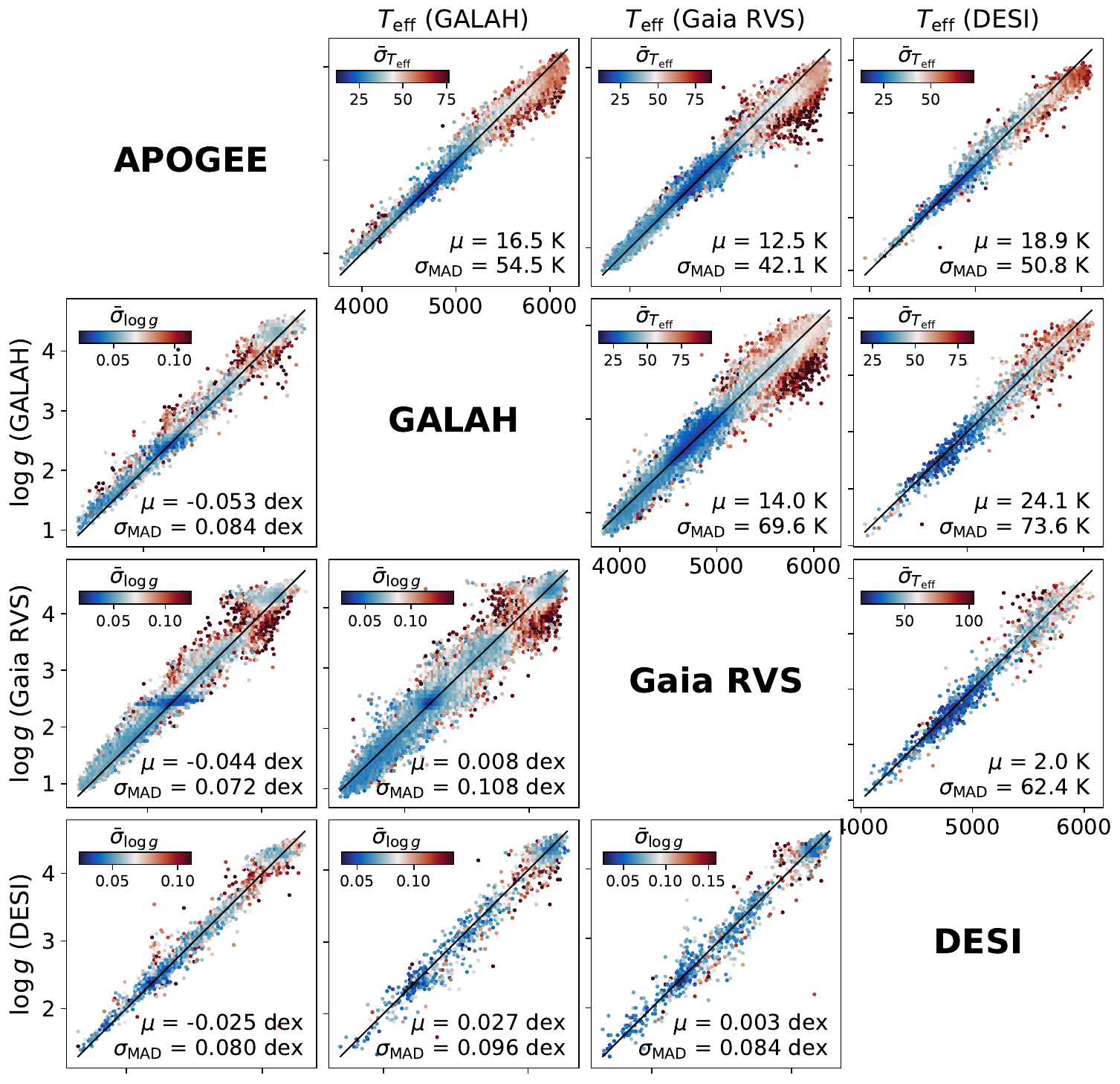}
  \caption{Cross-survey consistency check. Each off-diagonal panel
    compares predicted labels for stars observed by two different
    surveys. Panels above/to the right of the diagonal show $\teff$,
    and panels below/to the left show $\logg$. The diagonal labels
    indicate the corresponding survey. All panels are coloured by the
  larger of the two model-predicted uncertainties.}
  \label{fig:cross_survey_comparison}
\end{figure*}

In Figure \ref{fig:cross_survey_comparison} we show predicted
atmospheric parameters from one survey against those from another survey.
The diagonal indicates which survey that row/column corresponds to.
Above/to the right of the diagonal, we show the predicted values for
$\teff$, and below/to the left of the diagonal, we show the predicted
values for $\logg$.
All panels are coloured by the model prediction error, taking the
error that is larger between the two surveys (although the results
  look similar regardless of which survey the error is taken from, or
even if we take the average of the errors between the two surveys).
Despite some scatter in the relations, the points are fairly tightly
clustered around the 1-to-1 line, with no non-linearities or
otherwise strong trends in the residuals.
Furthermore, the model uncertainties still appear to be reasonably
well calibrated: errors are lower for points that agree, and larger
for points that disagree.
The exception is for stars with $\logg \sim 4$, where the model flags
even predictions that agree between surveys as unreliable.

Promisingly, the scatter in the residuals between the predictions
from the different surveys is generally on the order of, or slightly
smaller than, the residual in the internal validation of the
lower-quality survey alone (see Table
\ref{tab:performance-by-survey}), indicating cross-survey agreement
to the level of, or below, typical uncertainties.
Furthermore, the means of the residuals (i.e., systematic offsets)
are small, typically less than $20~$K for $\teff$ and less than
$0.05~$dex for $\logg$.
This check gives us confidence that there are no inter-survey
systematics at the label level, and that the labels have indeed been
homogenized across surveys.

\begin{figure*}[t]
  \centering
  \includegraphics[width=0.9\linewidth]{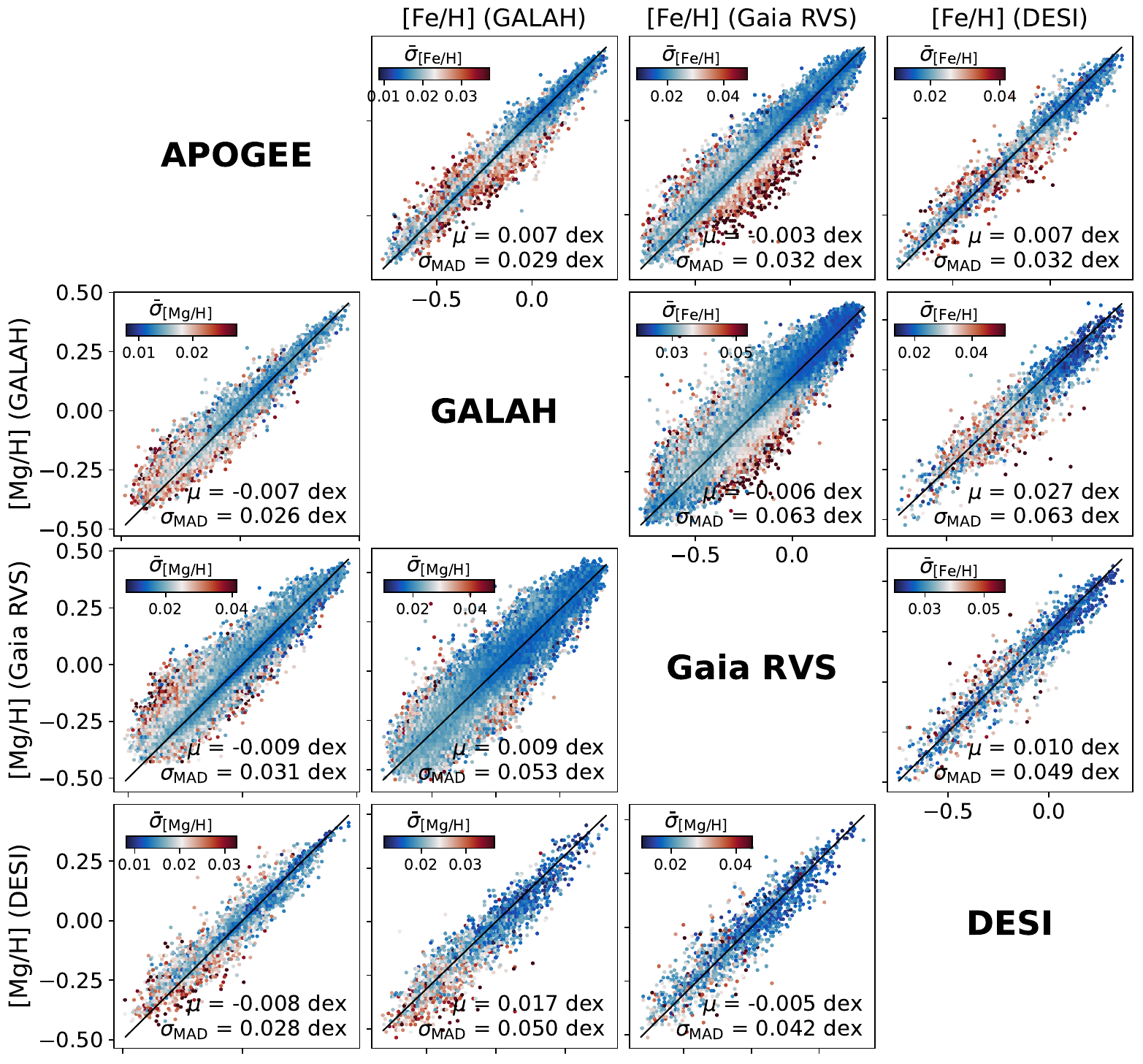}
  \caption{Same as Figure \ref{fig:cross_survey_comparison}, but for
  [Fe/H] and [Mg/Fe].}
  \label{fig:cross_survey_comparison_metallicity}
\end{figure*}

We show a similar check for [Fe/H] and [Mg/H] in Figure
\ref{fig:cross_survey_comparison_metallicity};
again we see that the means and scatters in the residuals are both small.
In particular, the mean in the residual in both iron abundance and
magnesium abundance is typically less than $0.01~$dex, indicating the
potential to make measurements of Galactic structure trends in
metallicity with high precision without the worry of false signals
due to systematic offsets.
Again the scatter in the residuals is typically on the order of the
measurement uncertainties, indicating that the cross-survey
measurements have been homogenized.

\subsection{Open clusters}

Open clusters are among the most valuable benchmarks in stellar astrophysics.
Because their member stars formed from the same molecular cloud at
approximately the same time, they share a common distance, age, and
initial chemical composition \citep[e.g.,][]{Bovy2016homogeneity}.
This chemical homogeneity makes open clusters an ideal testbed for
validating spectroscopic pipelines: if our model is performing well,
per-star label estimates for members of a given cluster should agree
with each other (low intra-cluster scatter) and also agree with
independent literature determinations of the cluster's metallicity and age.

We use the \citet{cantat-gaudin_gaia_2018} and
\citet{Spina2021cluster} catalogs to determine cluster membership.
We join the two catalogs and drop any duplicates based on the
\emph{Gaia} DR2 ID.
This leaves us with 2390 matches within a 1-arcsecond radius when we
cross-match with our catalog.

\subsubsection{Metallicities}

We use the metallicity estimates for open clusters from
\citet{Heiter2014cluster}, \citet{carrera2019cluster},
\citet{Dias2021cluster}, and \citet{Spina2021cluster}.
As in \citet{Spina2021cluster}, we remove clusters with median $\feh
< -0.5$, which are potentially unreliable.
We filter our cross-match to keep only stars with membership
confidence \texttt{Pmem > 0.5}, and further restrict to clusters with
more than 5 members in our catalog, leaving 48 clusters.

\begin{figure*}
  \centering
  \includegraphics[width=1\linewidth]{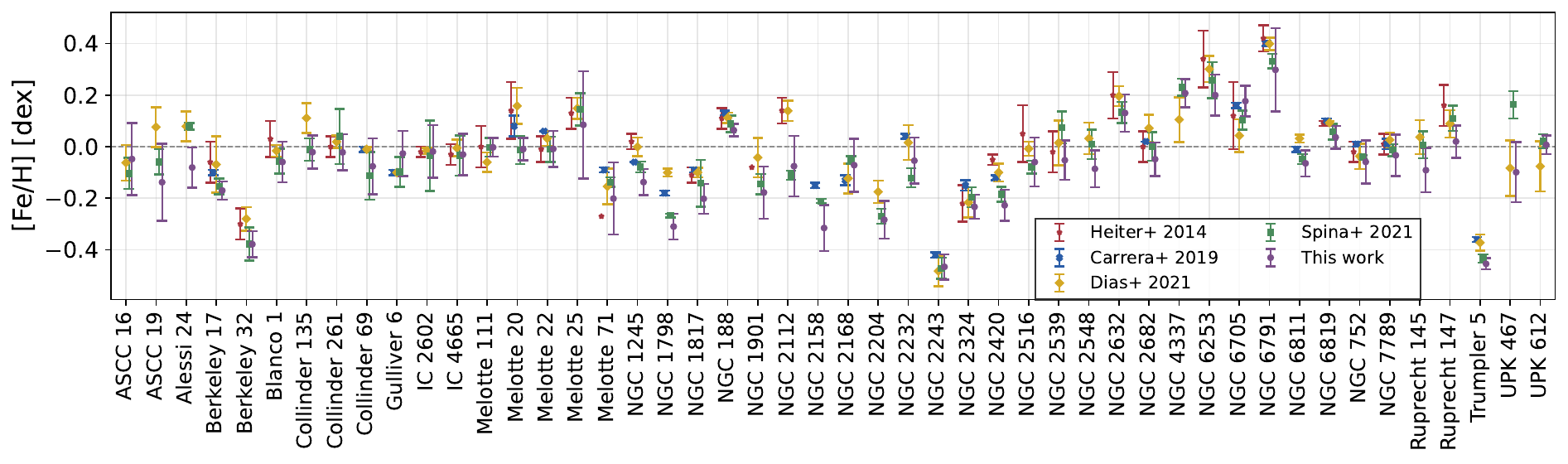}
  \caption{Comparison of cluster metallicities from
    \citet{Heiter2014cluster}, \citet{carrera2019cluster},
    \citet{Spina2021cluster}, and \citet{Dias2021cluster} to the median
    $\feh$ estimated from our catalog for member stars in each cluster.
    Error bars on our estimates indicate the standard deviation of our
    per-star $\feh$ estimates within each cluster. Only clusters with
  more than 5 members in the cross-match with our catalog are shown.}
  \label{fig:cluster_feh}
\end{figure*}

In Figure \ref{fig:cluster_feh}, we compare the $\feh$ values from
literature to those estimated from our catalog.
For each cluster, we take the median of the predicted $\feh$ across
all member stars as our estimate, with the standard deviation as the errorbar.
We find good agreement between our estimates and the literature values.
For the vast majority of clusters, our estimated $\feh$ agree with at
least one literature measurement to less than $1\sigma$.
Notably, the scatter, or difference between our measurements and
literature measurements, is no larger than the scatter between the
literature measurements themselves.
We calculate $\sigma_{\rm MAD}$ across all clusters between the
estimates for all combinations of pairs of literature measurements
from \citet{Heiter2014cluster}, \citet{carrera2019cluster},
\citet{Dias2021cluster}, and \citet{Spina2021cluster} (that is,
  $\sigma_{\rm MAD}$ between the \citet{Heiter2014cluster} and
  \citet{carrera2019cluster} estimates, between the
  \citet{Heiter2014cluster} and \citet{Spina2021cluster} estimates,
  between the \citet{Dias2021cluster} and \citet{Spina2021cluster}
estimates, and so on).
We also calculate $\sigma_{\rm MAD}$ between our estimates and each
of the literature measurements.
We then take the mean over all these estimated $\sigma_{\rm MAD}$
values: for both the intra-literature measurement and for the
measurement between our estimates and literature estimates, this
value is $0.05$~dex.
This provides strong confirmation that the metallicities in our
catalog are reliable and consistent with independent measurements.

\subsubsection{Distances}

We validate our distance estimates externally against three
independent catalogs:
(1) the spectrophotometric parallaxes of \cite{Hogg2019distance},
derived from APOGEE spectra and multi-band photometry;
(2) the DESI MWS SpecDis value-added catalog of \cite{Li2025specdis}; and
(3) the \cite{Queiroz2023starhorse} StarHorse catalog, which uses a
Bayesian isochrone-fitting approach to estimate distances for a
number of spectroscopic surveys, including APOGEE, GALAH, and \textit{Gaia} RVS.

We cross-match our catalog with each of these three references,
retaining only sources whose fractional distance uncertainty is less
than $7.5\%$ in both our catalog and the reference catalog.
For \cite{Hogg2019distance}, we match on 2MASS ID and compare against
their inverted parallaxes.
For \cite{Li2025specdis}, we match on sky coordinates within a
1-arcsecond radius, with no restriction to DESI-observed stars in our catalog.
For \cite{Queiroz2023starhorse}, we match each of their
survey-specific sub-catalogs (APOGEE, GALAH, \textit{Gaia} RVS) with
the corresponding survey in our catalog using the appropriate object
IDs, and aggregate the three cross-matched samples.
Because StarHorse reports asymmetric credible intervals rather than a
single uncertainty, we define the fractional distance uncertainty as
\begin{align}
  \frac{\sigma_{\rm dist}}{{\rm dist}_{50}} = \frac{({\rm dist}_{84}
  - {\rm dist}_{50}) + ({\rm dist}_{50} - {\rm dist}_{16})}{2 \, {\rm
  dist}_{50}},
\end{align}
where ${\rm dist}_{X}$ is the $X$-th percentile of the posterior distance.

\begin{figure*}
  \centering
  \includegraphics[width=1\linewidth]{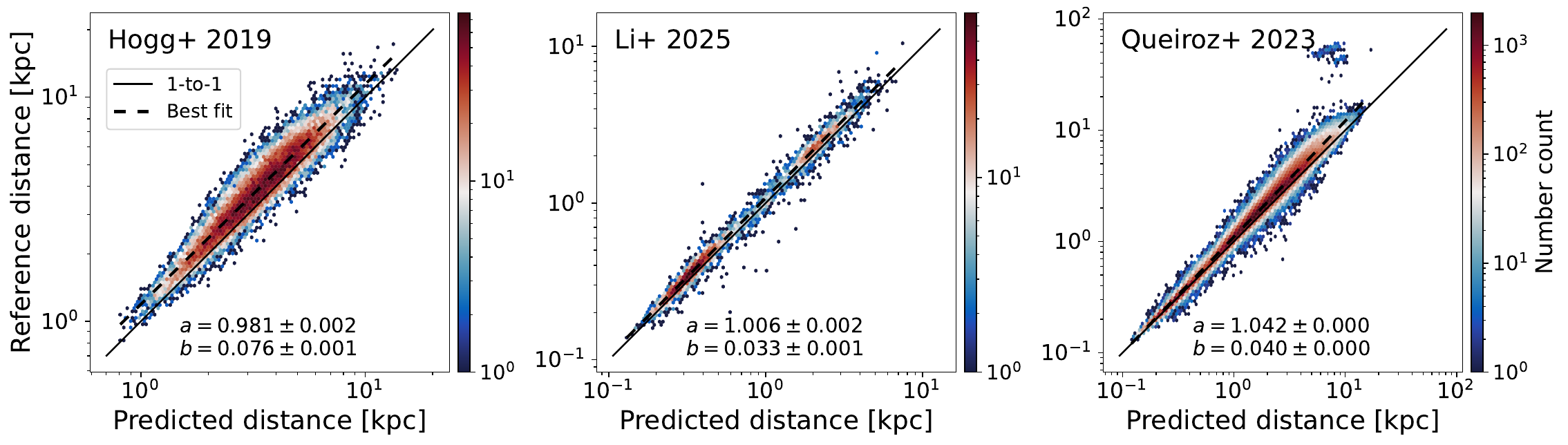}
  \caption{Comparison of our predicted distances to three external
    catalogs: \citet{Hogg2019distance} (left), \citet{Li2025specdis}
    (centre), and \citet{Queiroz2023starhorse} (right). Each panel shows
    a hexbin density plot of reference distance vs. predicted distance in
    log--log space, coloured by number count. The solid line marks the
    1-to-1 relation, and the dashed line is the best fit linear model
    $\log d_{\rm ref} = a\,\log d_{\rm pred} + b$, with parameters
    annotated in each panel. Only sources with fractional distance
  uncertainty less than $7.5\%$ in both catalogs are shown.}
  \label{fig:distance_comparison}
\end{figure*}

Figure~\ref{fig:distance_comparison} compares our distance estimates
to those from each of the three external catalogs.
In each panel, we plot our predicted distance on the $x$-axis and the
reference distance on the $y$-axis, both in log--log space, and fit a
linear relation $\log d_{\rm ref} = a\,\log d_{\rm pred} + b$.
Here, the slope $a$ measures whether our distances track the
reference distances proportionally across the full distance range,
with $a = 1$ implying no distance-dependent bias and any deviations
from unity indicating that the scaling between our distances and the
reference is distance-dependent.
The intercept $b$ captures any constant multiplicative offsets, since
$d_{\rm ref} = 10^{b}\,d_{\rm pred}^{a}$.

All three comparisons yield slopes close to unity ($a =
  0.981\pm0.002$, $1.006\pm0.002$, and $1.042\pm0.000$ for
  \citealt{Hogg2019distance}, \citealt{Li2025specdis}, and
\citealt{Queiroz2023starhorse}, respectively), indicating that our
distances track the reference distances proportionally across the
compared range (${\sim}10$~kpc) with no significant distance-dependent bias.
The tightest agreement is with \cite{Li2025specdis}, which shows both
the smallest scatter and a slope that is consistent with $1$ within $3\sigma$.

All three comparisons also show a small but consistent positive
intercept ($b = 0.033$--$0.076$~dex), indicating that our distances
are systematically underpredicted by ${\sim}8$--$19\%$.
This is consistent with the internal validation in
Figure~\ref{fig:eval_all}, where we find a mean residual bias of
$0.034$~dex in log distance ($8\%$).
We provide the best-fit parameters in each panel of
Figure~\ref{fig:distance_comparison} so that users may apply a
post-hoc correction if desired to bring our predicted distances onto
the scale of a given reference catalog.

\subsubsection{Ages}

In a similar manner to the metallicities, we validate our age
estimates on open cluster ages.
Since stars in a cluster are roughly co-eval, the scatter in ages
between stars is expected to be small, and isochrone fitting can be
used to obtain an accurate age for the cluster.
Here, we are stricter on our selection of stars.
We keep only stars with $\teff < 5200$~K, $\logg < 3.3$ and
$\sigma_{\logg} < 0.1$, as our ages come exclusively from
asteroseismology of red giants; hence, the model has never seen a
reliable age outside this parameter range (and in particular for
dwarf stars), and thus the age prediction is unreliable.
We also perform an age uncertainty cut, keeping only predicted ages
with an uncertainty of less than $30\%$, mirroring the cut used for
the training labels.
We also tighten the membership probability cut to keep only stars
with \texttt{Pmem > 0.9}.
Again we keep only clusters with at least 5 members, leaving only 22 clusters.

\begin{figure*}
  \centering
  \includegraphics[width=1\linewidth]{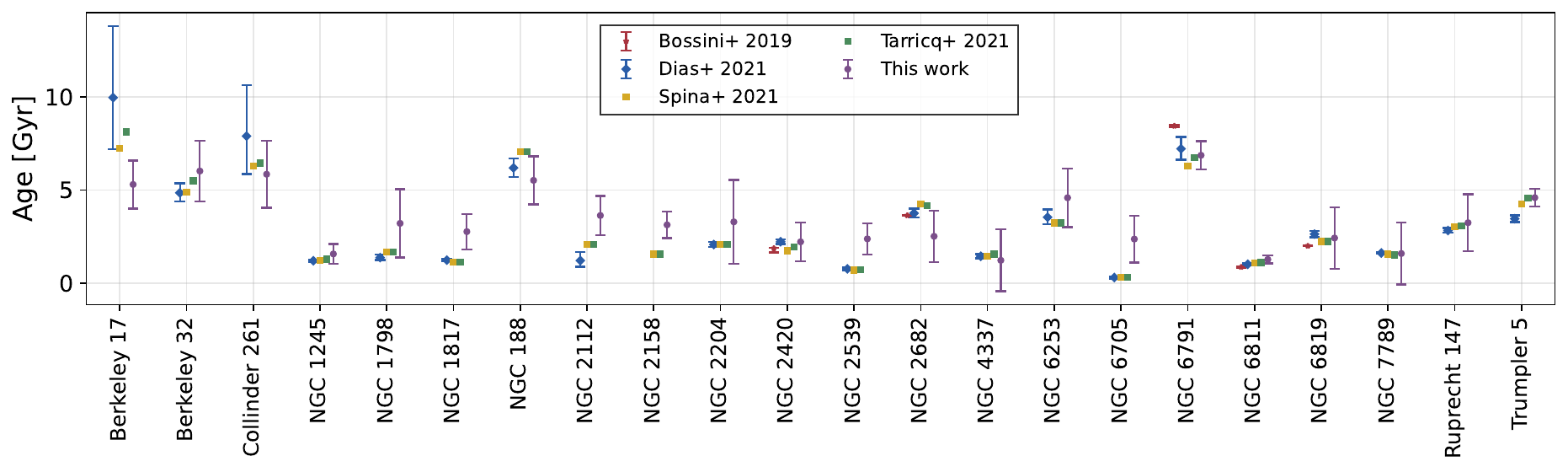}
  \caption{Comparison of cluster ages from \citet{bossini2019cluster},
    \citet{Dias2021cluster}, \citet{Spina2021cluster}, and
    \citet{tarricq2021cluster} to the median age estimated from our
    catalog for member stars in each cluster. Errorbars on our estimates
    indicate the standard deviation of our per-star age estimates within
    each cluster. Only clusters with more than 5 members in the
    cross-match with our catalog are shown. No errors are provided by
  \citet{Spina2021cluster} and \citet{tarricq2021cluster}.}
  \label{fig:cluster_age}
\end{figure*}

In Figure \ref{fig:cluster_age} we show our estimated ages for these
22 clusters against literature measurements.
Visually, there is less agreement between our estimates and
literature measurements than there is for our metallicity estimates.
It appears that our estimates are overestimated for young clusters,
and underestimated for old clusters; the latter is a known problem
with AstroNN ages \citep{Anders2023ages, Leung_2023variational}.
For many clusters the scatter is also high, indicating that estimates
for different stars in the same cluster disagree.
However, when the scatter is low, our estimates generally are in
agreement with literature measurements (e.g., NGC 1245, NGC 6811, Trumpler 5).

\section{Discussion and Conclusion}

\subsection{Choice of training labels}

A natural question is why we train on AstroNN labels---themselves the
output of a neural network \citep{leung_deep_2018}---rather than
directly on the ASPCAP pipeline labels from which they were derived.
Our choice can be viewed as an application of \emph{knowledge
distillation} \citep{Hinton2015distillation}, in which a student
model learns to reproduce the outputs of a teacher model rather than
the raw training targets.
Knowledge distillation has been widely successful across machine
learning, particularly in the training of compact large language
models (e.g., DistilBERT \citealt{sanh2020distilbert}, Gemma~2
\citealt{gemma2}).
Student models trained on teacher outputs often outperform students
trained directly on noisy ground-truth labels, because the teacher's
predictions are smoother and better calibrated
\citep[e.g.,][]{Hinton2015distillation, tang2021kd, Gou_2021}.

In our case, the AstroNN labels can be viewed as a smoothed, denoised
version of the ASPCAP labels.
ASPCAP abundances, while carefully calibrated, are derived from
$\chi^2$ fitting of synthetic spectra to observed spectra, and
individual measurements can be noisy or exhibit outlier behavior,
particularly for elements with weak spectral features or in low
signal-to-noise regimes.
The AstroNN model, trained on these same ASPCAP labels, learns the
underlying mapping from spectra to labels and in doing so implicitly
averages over the noise in individual ASPCAP measurements.
The resulting AstroNN labels are therefore less noisy on a
star-by-star basis \citep{leung_deep_2018} and present a more
learnable target for our model.
Training on these smoother labels can improve the precision and
stability of our predictions, analogous to the label smoothing effect
observed in knowledge distillation \citep{tang2021kd}.

\subsection{Limitations}

The main limitation of our framework is that it inherits the
systematics of the reference labels on which it is trained.
Because we use AstroNN labels---themselves derived from ASPCAP
abundances and asteroseismic ages---any systematic biases present in
those labels propagate into our predictions.
This is particularly relevant for stellar ages: the AstroNN ages are
trained on asteroseismic ages from APOKASC-2, and the AstroNN model
is known to systematically overestimate young ages and underestimate
old ages \citep{Anders2023ages, Leung_2023variational}, a clamping
effect that our model inherits (see Figure~\ref{fig:cluster_age}).
More broadly, any element for which ASPCAP abundances are
unreliable---such as [Co/H], [Na/H], and [P/H]
\citep{leung_deep_2018}---will also be less reliable in our catalog.
This is, however, an inherent limitation of any supervised machine
learning approach that relies on a training set; the key advantage of
our framework is that it avoids the \emph{inter-survey} systematics
that arise when each survey is analyzed with an independent pipeline.

We further caution that ``homogeneous'' should not be interpeted as
``accurate in an absolute sense.''
What our framework is designed to achieve is that all surveys are
placed on a single, \emph{self-consistent} scale---concretely here,
the APOGEE/AstroNN scale, but more generally for this framework
whichever set of reference labels if chosen---so that any systematic
biases present in the reference labels is shared equally across all
input surveys.
In other words, homogenization eliminates relative systematic offsets
between surveys, but it does not correct for any global biases which
are inherited from the reference labels.
For example, if AstroNN systematically overestimates [Fe/H], our
predicted [Fe/H] for every survey will be overestimated by the same amount.
This means that catalogs produced by this framework are best suited
for analyses where internal consistency matters more than a correct
absolute calibration (e.g., measuring abundance gradients,
  identifying chemical substructure, or comparing stellar populations
across surveys).

The training label distribution also limits the parameter range over
which predictions are reliable.
In particular, the AstroNN sample is dominated by red giants with few
low-metallicity stars: less than $10\%$ of the training sample has
$\feh < -0.5$, and there are essentially no stars with $\feh < -1$
after quality cuts.
Predictions in the metal-poor regime are therefore less reliable and
often subject to edge effects, as reflected in the
out-of-distribution flagging described in Section~\ref{subsec:flagging}.
This will limit the use of our catalog for studies of, for example,
the metal-poor halo and accreted substructures.
Similarly, our age estimates are only reliable for red giant stars,
because the training ages are derived exclusively from
asteroseismology of \emph{Kepler} red giants \citep{Pinsonneault2018}.
The model has not been exposed to reliable age information for
dwarfs, and age predictions outside the giant branch should not be used.

A further limitation is that we do not explicitly account for
interstellar extinction in our analysis.
Interstellar dust attenuates and reddens stellar spectra in a way
that depends on the line-of-sight dust column and the
wavelength-dependent extinction law $A(\lambda)$, and in principle it
would be more rigorous to have an extinction-aware model.
For example, one could do this by providing an estimate of $A_V$ from
a 3D dust map as an additional input.
Including this additional information could possibly deliver more
accurate predictions, particularly in the most heavily reddened
regions such as the inner disk and bulge, and particularly for
spectrophotometric distances, where extinction is directly tied to
the apparent flux scale.
Nevertheless, in this work we choose not to explicitly model
extinction, for several reasons that we outline below.

First, while we do not model extinction explicitly, the structure of
our input pipeline reduces its impact on our predictions.
Before patching, each spectrum is normalized by its median flux,
which factors the input into two pieces: a \emph{shape} component
(the median-normalized flux pattern, processed by the encoder as a
sequence of patches) and a \emph{scale} component (the median itself,
  which is preserved as a single additional token and passed through
the regression head).
Extinction affects these two components very differently.
The median flux is strongly attenuated by extinction; consequently,
spectrophotometric distances, which rely on the absolute flux scale,
are most affected.
The median-normalized spectrum, however, is much less sensitive to
extinction, because extinction acts approximately as a smooth
multiplicative modulation of the flux as a function of wavelength,
and normalizing by the median removes its overall level.
What remains after normalization is a residual wavelength-dependent
tilt that follows the shape of the extinction curve across the
observed wavelength window.

While $A(\lambda)$ is a smooth function that varies on large
wavelength scales (roughly $\propto 1/\lambda$ in the optical and
even flatter in the near-infrared), the information used to constrain
atmospheric parameters and chemical abundances lives in spectral line
features with characteristic widths of order ${\sim}\,0.1$--$1$~\AA.
Extinction and stellar lines therefore occupy different scales
wavelength: extinction is a low-frequency modulation of the continuum
that is captured well by a low-order polynomial, whereas the line
features that contain information about the abundances live at much
higher frequencies that $A(\lambda)$ does not affect strongly.
In fact, spectroscopic pipelines such as ASPCAP also operate on
pseudo-continuum normalized spectra, removing variations in spectral
shape due to reddening in their fits for atmospheric parameters and abundances.
We note, however, that our model does not explicitly separate
spectral lines from continuum, and if the continuum tilt happens to
correlate with a label in the training data, the model is able to,
and will likely exploit the tilt to improve the label estimation,
which could possibly introduce a residual extinction-dependent bias.

Second, the reference labels we train against already reduce the
effect of extinction that our model might otherwise need to handle itself.
The AstroNN distances use extinction-corrected apparent magnitudes
$K_{s,0}$ to convert predicted spectrophotometric luminosities into a
spectrophotometric parallax \citep{mackereth_dynamical_2019}.
Since the network predicts the luminosity from continuum-normalized
spectra, the effect of reddening is once again mitigated and limited
to a residual effect dependent on the imperfections of the continuum
normalization.
Our model, trained on AstroNN labels, inherits its extinction-reduced
scale rather than having to recover it entirely from scratch.
The same reasoning applies, though to a lesser extent, to the
atmospheric parameters and abundances.
The AstroNN atmospheric parameters and abundances are ultimately
derived from ASPCAP, which operates on pseudo-continuum normalized
APOGEE spectra;
the bulk of the smooth extinction-induced tilt is therefore already
removed before any fitting of these labels takes place, and the
reddening signal that propagates into AstroNN (and subsequently into
our model) is correspondingly suppressed.

Third, the properties of the APOGEE survey itself are favourable for
an approach that does not model extinction explicitly, in two
complementary ways.
On the one hand, APOGEE observes in the near-infrared
($1.51$--$1.70~\mu\text{m}$), where extinction is weaker than in the
optical by roughly an order of magnitude ($A_H / A_V \sim 0.15$);
since the reference labels used in training were extracted from
APOGEE spectra, even stars with substantial optical reddening present
only a much reduced extinction signal to the pipelines that generated
our training labels, and any extinction-induced systematics in the
training set are correspondingly smaller in absolute terms than one
would expect from an optical-based reference.
On the other hand, APOGEE was deliberately designed to target
high-extinction regions of the disk and bulge that are largely
inaccessible to optical surveys \citep{Majewski2017}, so the training
sample spans a broad range of line-of-sight extinctions, from
essentially unreddened halo stars to heavily embedded disk and bulge giants.
The two facts compound: the training labels sit on a relatively
extinction-insensitive (near-infrared) scale to begin with, and our
model is exposed during training to the full variation of extinction
conditions represented by APOGEE and must learn to predict the same
(extinction-corrected) AstroNN labels across that range.
This gives the model the opportunity to develop some robustness to
extinction from the data alone, without being told about it explicitly.

We note that the cross-survey consistency check in
Section~\ref{subsec:cross-survey} provides a further, empirical
constraint: if unmodelled extinction were strongly biasing our
labels, we would expect stars observed by multiple surveys with
different wavelength coverages---and therefore different extinction
sensitivities---to disagree systematically.
We do not observe such disagreements at levels above our quoted
precisions, which places an informal upper limit on how much residual
extinction-dependent bias could be present in the current catalog.
Even so, we stress that none of the arguments above amount to a full
solution, and a properly extinction-aware treatment---whether via an
explicit extinction input, joint marginalization, or an
extinction-augmented training set---remains a worthwhile direction
for future work, particularly for extending the catalog into the most
heavily reddened regions of the Galaxy and for improving the
reliability of spectrophotometric distances along high-extinction sightlines.

\subsection{Summary}

We have presented a unified deep-learning framework for deriving
homogeneous stellar labels from heterogeneous spectroscopic surveys.
Building on the spectral tokenizer architecture of
\citet{Shen2025tokenizer}, we train a single Transformer-based
regression model that ingests spectra of arbitrary wavelength range
and resolution, directly and simultaneously producing atmospheric
parameters ($\teff$, $\logg$), chemical abundances for 20 elements,
spectrophotometric distances, and stellar ages on a single,
self-consistent scale.

We apply this framework to spectra from four major surveys with
low-resolution and high-resolution optical and near-IR spectra.
The resulting catalog is accompanied by label uncertainties derived
from Monte Carlo Dropout, as well as out-of-distribution detection
flags based on Mahalanobis distances, Isolation Forest scores, and
GMM likelihoods, applied to the model's internal embeddings, enabling
flexible downstream quality filtering.

Internal validation on a held-out set demonstrates excellent
precision: on APOGEE spectra, the model achieves $18$~K in $\teff$,
$0.04$~dex in $\logg$, $0.015$~dex in $\feh$, and ${<}\,0.03$~dex
across all chemical abundances; on lower-resolution surveys such as
DESI, typical precisions are $51$~K, $0.09$~dex, $0.04$~dex, and
${\sim}\,0.06$~dex, respectively.
The predicted uncertainties are positively correlated with actual
residuals, providing users a reliable and straightforward means of
filtering out less trustworthy estimates.
Cross-survey validation confirms that labels predicted for the same
star from different surveys agree to within model uncertainties, with
no strong evidence of cross-survey systematics, demonstrating that
the labels have indeed been homogenized.
External validation against open cluster metallicities from the
literature shows good agreement, with our per-cluster scatter
comparable to the scatter among literature measurements themselves.
Age estimates, however, show larger scatter and systematic trends,
highlighting a direction for future improvement.

\subsection{Outlook}

The framework presented here is readily extensible to forthcoming
large-scale spectroscopic surveys.
Because our architecture is agnostic to spectral resolution and
wavelength coverage, incorporating new surveys requires only adding
them to the training set, with no architectural modifications.
In this way, our work moves in the same direction as other recent
work on multimodal foundation models for astronomy
\citep{Parker2025aion}, which have demonstrated the potential of
unified architectures that perform large-scale pretraining on
heterogeneous data from multiple surveys simultaneously.
We plan to release an updated model pre-trained on a larger and more
diverse set of spectroscopic surveys, as well as a set of labels with
broader parameter coverage, with the aim to improve predictions in
currently under-sampled regimes.

Beyond extending the survey coverage, several natural extensions of
this work are worth pursuing.
The cross-attention pooling mechanism is already agnostic to the
length of the input token sequence, meaning that multiple spectra of
the same star---potentially from different surveys---could be
combined into a single, more tightly constrained set of labels.
Expanding the training label set to include ages from subgiant and
main-sequence stars, as well as more metal-poor reference stars,
would substantially broaden the regime over which reliable
predictions can be made.

Another direction is to investigate where in the spectrum the label
information originates.
Gradient-based attribution methods and related techniques
\citep{Simonyan2014saliency, Sundararajan2017ig,
Smilkov2017smoothgrad, Shen2023attribution}---have become standard
tools for interpreting deep neural networks and can be applied to our
model to produce per-label gradient,
$\partial\,\mathrm{label}/\partial\,\mathrm{flux}(\lambda)$, that
highlight the wavelength regions most responsible for a given prediction.
Such analyses have been applied to the interpretation of data-driven
stellar label models to verify that predicted abundances are
sensitive to physically meaningful spectral features
\citep{leung_deep_2018, Fabbro2018, OBriain2021cyclestarnet} and to
connect learned sensitivities to Cramer--Rao information bounds on
chemical abundance precision \citep{Ting2017prospects, Sandford2020}.
Extending this to our multi-survey framework would both validate that
the model relies on physically meaningful features and to test
whether the same wavelength regions carry the same label information
across surveys with different resolutions and wavelength coverage.

In a companion paper (Shen et al., in prep.), we will perform
chemical cartography of the Milky Way, exploiting the homogeneity of
the labels to map abundance gradients across this unified multi-survey sample.

\begin{acknowledgments}

  J.S. is supported by the Natural Sciences and Engineering Research
  Council of Canada (NSERC), funding reference number 587652, and by
  the Citadel GQS PhD Fellowship in Physics.
  J.S.S. was supported by funding from NSERC Discovery Grant RGPIN-2023-04849.
  The computations in this work were carried out on, and the data
  hosted on equipment supported by the Scientific Computing Core at the
  Flatiron Institute, a division of the Simons Foundation.

  This work made use of the Third Data Release of the GALAH Survey
  (Buder et al. 2021). The GALAH Survey is based on data acquired
  through the Australian Astronomical Observatory, under programs:
  A/2013B/13 (The GALAH pilot survey); A/2014A/25, A/2015A/19,
  A2017A/18 (The GALAH survey phase 1); A2018A/18 (Open clusters with
  HERMES); A2019A/1 (Hierarchical star formation in Ori OB1); A2019A/15
  (The GALAH survey phase 2); A/2015B/19, A/2016A/22, A/2016B/10,
  A/2017B/16, A/2018B/15 (The HERMES-TESS program); and A/2015A/3,
  A/2015B/1, A/2015B/19, A/2016A/22, A/2016B/12, A/2017A/14 (The HERMES
  K2-follow-up program). We acknowledge the traditional owners of the
  land on which the AAT stands, the Gamilaraay people, and pay our
  respects to elders past and present. This paper includes data that
  has been provided by AAO Data Central (datacentral.org.au).

  This work has made use of data from the European Space Agency (ESA) mission
  {\it Gaia} (\url{https://www.cosmos.esa.int/gaia}), processed by
  the {\it Gaia}
  Data Processing and Analysis Consortium (DPAC,
  \url{https://www.cosmos.esa.int/web/gaia/dpac/consortium}). Funding
  for the DPAC
  has been provided by national institutions, in particular the institutions
  participating in the {\it Gaia} Multilateral Agreement.

  This research used data obtained with the Dark Energy Spectroscopic
  Instrument (DESI). DESI construction and operations is managed by the
  Lawrence Berkeley National Laboratory. This material is based upon
  work supported by the U.S. Department of Energy, Office of Science,
  Office of High-Energy Physics, under Contract No. DE–AC02–05CH11231,
  and by the National Energy Research Scientific Computing Center, a
  DOE Office of Science User Facility under the same contract.
  Additional support for DESI was provided by the U.S. National Science
  Foundation (NSF), Division of Astronomical Sciences under Contract
  No. AST-0950945 to the NSF’s National Optical-Infrared Astronomy
  Research Laboratory; the Science and Technology Facilities Council of
  the United Kingdom; the Gordon and Betty Moore Foundation; the
  Heising-Simons Foundation; the French Alternative Energies and Atomic
  Energy Commission (CEA); the National Council of Humanities, Science
  and Technology of Mexico (CONAHCYT); the Ministry of Science and
  Innovation of Spain (MICINN), and by the DESI Member Institutions:
  www.desi.lbl.gov/collaborating-institutions. The DESI collaboration
  is honored to be permitted to conduct scientific research on I’oligam
  Du’ag (Kitt Peak), a mountain with particular significance to the
  Tohono O’odham Nation. Any opinions, findings, and conclusions or
  recommendations expressed in this material are those of the author(s)
  and do not necessarily reflect the views of the U.S. National Science
  Foundation, the U.S. Department of Energy, or any of the listed
  funding agencies.

  Funding for the Sloan Digital Sky
  Survey IV has been provided by the
  Alfred P. Sloan Foundation, the U.S.
  Department of Energy Office of
  Science, and the Participating
  Institutions.

  SDSS-IV acknowledges support and
  resources from the Center for High
  Performance Computing at the
  University of Utah. The SDSS
  website is www.sdss4.org.

  SDSS-IV is managed by the
  Astrophysical Research Consortium
  for the Participating Institutions
  of the SDSS Collaboration including
  the Brazilian Participation Group,
  the Carnegie Institution for Science,
  Carnegie Mellon University, Center for
  Astrophysics | Harvard \&
  Smithsonian, the Chilean Participation
  Group, the French Participation Group,
  Instituto de Astrof\'isica de
  Canarias, The Johns Hopkins
  University, Kavli Institute for the
  Physics and Mathematics of the
  Universe (IPMU) / University of
  Tokyo, the Korean Participation Group,
  Lawrence Berkeley National Laboratory,
  Leibniz Institut f\"ur Astrophysik
  Potsdam (AIP),  Max-Planck-Institut
  f\"ur Astronomie (MPIA Heidelberg),
  Max-Planck-Institut f\"ur
  Astrophysik (MPA Garching),
  Max-Planck-Institut f\"ur
  Extraterrestrische Physik (MPE),
  National Astronomical Observatories of
  China, New Mexico State University,
  New York University, University of
  Notre Dame, Observat\'ario
  Nacional / MCTI, The Ohio State
  University, Pennsylvania State
  University, Shanghai
  Astronomical Observatory, United
  Kingdom Participation Group,
  Universidad Nacional Aut\'onoma
  de M\'exico, University of Arizona,
  University of Colorado Boulder,
  University of Oxford, University of
  Portsmouth, University of Utah,
  University of Virginia, University
  of Washington, University of
  Wisconsin, Vanderbilt University,
  and Yale University.

\end{acknowledgments}

\begin{contribution}

  J.S. developed the idea, compiled the data, designed and trained the
  model, performed the analysis, and wrote the manuscript.
  J.S.S. contributed scientific expertise, contributed to the
  interpretation, and provided guidance on the presentation of this manuscript.
  S.H. acquired computing resources for the project and contributed to
  the interpretation.

\end{contribution}

\software{Astropy \citep{astropy:2013, astropy:2018, astropy:2022},
  Numpy \citep{harris2020array}, Matplotlib \citep{Hunter:2007},
  PyTorch \citep{paszke2019pytorchimperativestylehighperformance},
  scikit-learn \citep{scikit-learn}, cmocean \citep{thyng2016cmocean},
LanceDB \citep{pace2025lanceefficientrandomaccess}, DuckDB \citep{duckdb}}

\appendix

\section{Uncertainty calibration}

\begin{figure*}[ht]
  \centering
  \includegraphics[width=1\linewidth]{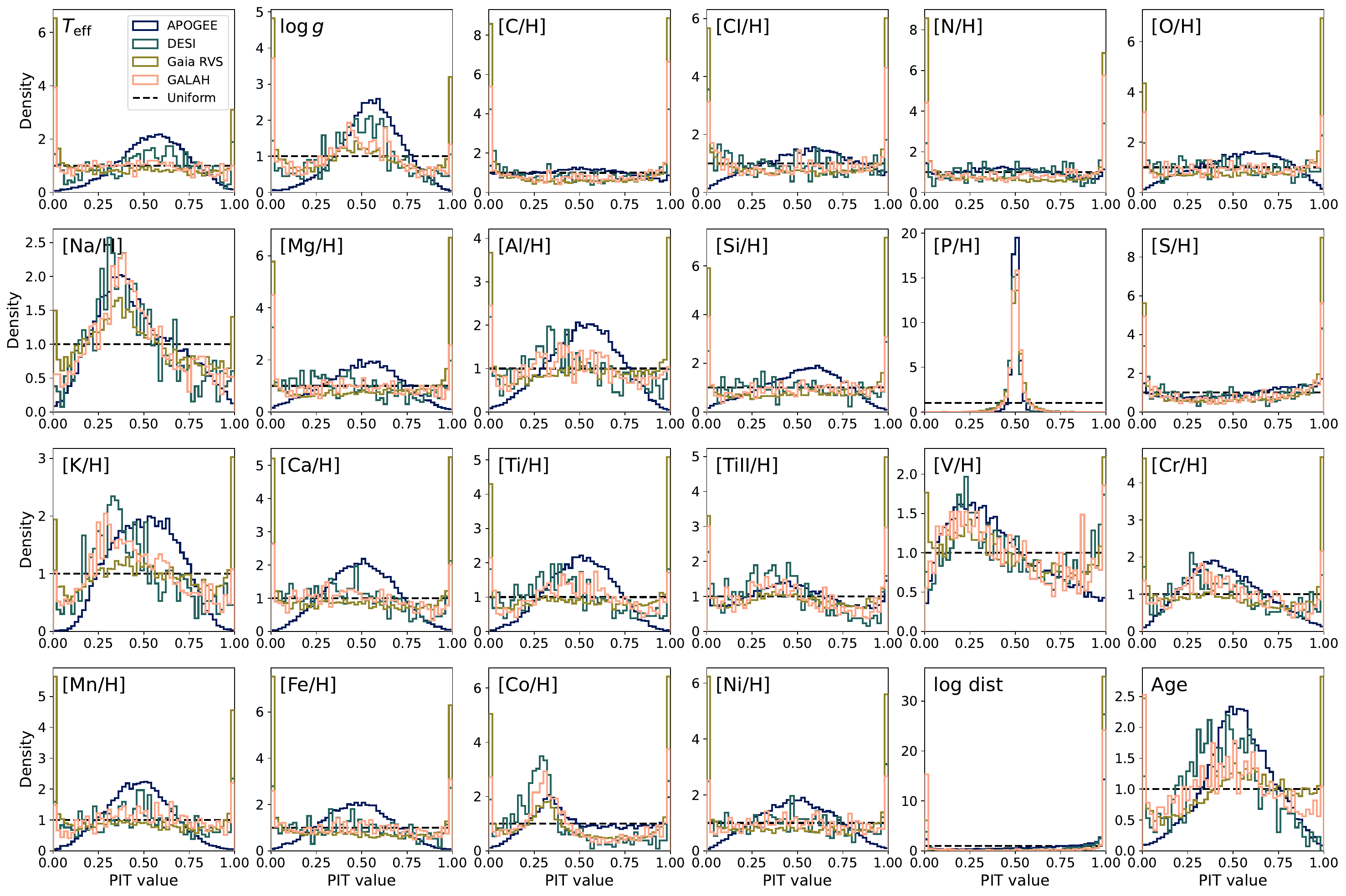}
  \caption{Same as Figure \ref{fig:calibration}, but calculated using
    only the label errors, and not the sum in quadrature of the
  predictive errors and the label errors.}
  \label{fig:calibration_labelerr_only}
\end{figure*}

We show in Figure \ref{fig:calibration_labelerr_only} PIT histograms
for all predicted labels, broken down by input survey, where the
z-score is calculated using only the reference survey's uncertainty estimates.
This gives an idea of the uncertainty floor that we are working with.

\section{Comparison with pipeline labels}

\begin{figure}[ht]
  \centering
  \includegraphics[width=1\linewidth]{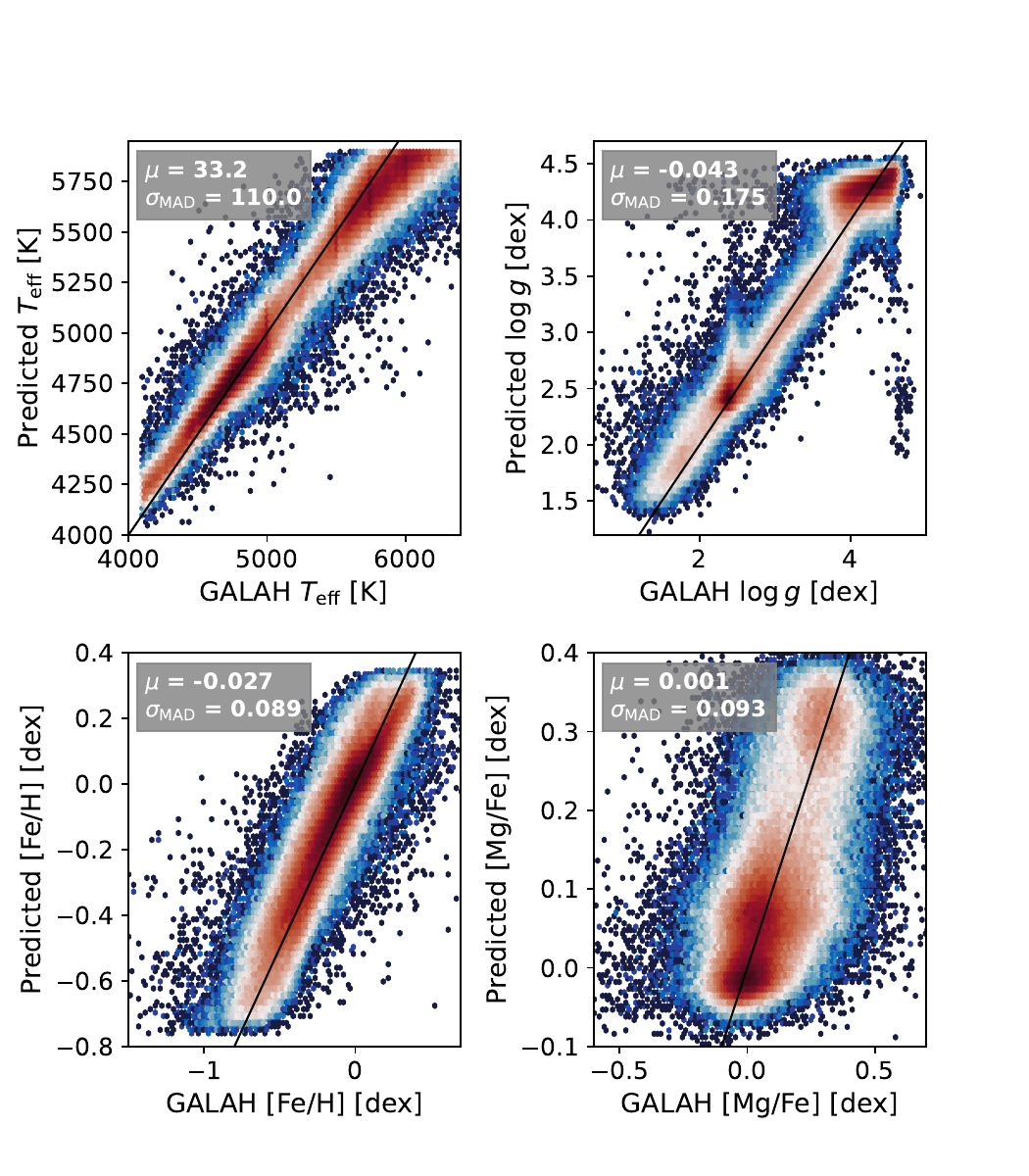}
  \caption{Comparison of our predicted stellar parameters from GALAH
    spectra to the values reported by the GALAH pipeline for $\teff$ (top
    left), $\logg$ (top right), $\feh$ (bottom left), and [Mg/Fe] (bottom
    right). Quality cuts have been applied to both sets of labels (see
    text). The black line indicates the 1-to-1 relation. Systematic
    differences between the two pipelines are visible, including a slope
  difference in $\teff$ and a small constant offset in $\feh$.}
  \label{fig:compare_to_galah}
\end{figure}

As an additional external check, we compare our predicted labels for
GALAH stars against the labels reported by the GALAH pipeline
(Figure~\ref{fig:compare_to_galah}).
We emphasize that this is a comparison against an \emph{independent}
set of reference labels: although the GALAH spectra themselves are
part of our training set, the GALAH pipeline labels are never seen by the model.
Our model is trained exclusively on AstroNN labels
\citep{leung_deep_2018}, which are derived from APOGEE spectra
through the ASPCAP pipeline;
the GALAH pipeline labels used here are derived independently via the
Spectroscopy Made Easy approach \citep{Piskunov2017sme,
buder_galah_2021}, based on optical spectra, with different synthetic
spectra, line lists, and microturbulence prescriptions.
The two reference systems are therefore independent, and this
comparison tests whether the labels our model predicts from GALAH
spectra are consistent with an external, independently derived set of
labels for the same stars.

Before interpreting the comparison, we stress that perfect agreement
with the GALAH pipeline is neither expected nor desired.
Systematic differences between APOGEE/ASPCAP and GALAH labels---in
$\teff$, $\logg$, $\feh$, and individual abundances---have been
extensively documented in the literature
\citep[e.g.,][]{buder_galah_2021, Nandakumar2022,
Jofre2019industrial}, and arise from fundamental differences in the
two pipelines' underlying models and calibration choices rather than
from errors in either.
Because our model is trained on APOGEE/AstroNN labels, we expect it
to reproduce the APOGEE-scale labels when applied to GALAH spectra,
and thus to exhibit the same APOGEE--GALAH offsets reported in these
prior studies.
The purpose of this comparison is therefore not to validate our
labels against GALAH, but rather (i) as a sanity check that the model
produces sensible predictions on GALAH spectra, and (ii) as a way to
identify and characterize the expected inter-pipeline systematics in
our catalog, so that downstream users can interpret our labels appropriately.

We apply several quality cuts to both sets of labels: on the GALAH
side, we require \texttt{snr\_c3\_iraf > 30}, \texttt{flag\_sp = 0},
and \texttt{flag\_fe\_h = 0}; on our side, we apply the same quality
cuts as in Section~\ref{subsec:cross-survey}, and additionally remove
predicted labels that fall below the 1st or above the 99th percentile
of the training distribution for that label, restricting the
comparison to a range where our predictions are reasonably reliable.

Overall, our predicted values track the GALAH pipeline values well,
with a few systematic differences that are broadly consistent with
the APOGEE--GALAH offsets reported in the literature.
For $\teff$, agreement is good near $\teff \sim 5100~{\rm K}$, but
diverges at higher and lower temperatures: GALAH predicts higher
$\teff$ than our model for hotter stars and lower $\teff$ for cooler
stars, consistent with a difference in slope between the two pipelines.
For $\logg$, our measurements agree well with GALAH, though we
slightly overestimate relative to GALAH at low surface gravities.
For $\feh$, we find a small but consistent offset
(${\sim}\,0.03$~dex) across the entire metallicity range.

\nocite{Shen2026_homogeneous_zenodo}
\bibliography{references}{}
\bibliographystyle{aasjournalv7}

\end{document}